\documentclass[pdflatex,sn-mathphys-num]{sn-jnl}

\usepackage{graphicx}%
\usepackage{multirow}%
\usepackage{amsmath,amssymb,amsfonts}%
\usepackage{amsthm}%
\usepackage{mathrsfs}%
\usepackage[title]{appendix}%
\usepackage{xcolor}%
\usepackage{textcomp}%
\usepackage{manyfoot}%
\usepackage{booktabs}%
\usepackage{algorithm}%
\usepackage{algorithmicx}%
\usepackage{algpseudocode}%
\usepackage{listings}%

\theoremstyle{thmstyleone}%
\theoremstyle{thmstyletwo}%

\theoremstyle{thmstylethree}%

\newcommand{\dint}{\mathrm{d}}

\begin{document}

\title[Extrapolation from historical data cannot reliably predict the time of a potential AMOC collapse]{Extrapolation from historical data cannot reliably predict the time of a potential AMOC collapse}

\author*[1,2]{\fnm{Andreas} \sur{Morr}}\email{andreas.morr@tum.de}
\author[2,3]{\fnm{Maya} \sur{Ben-Yami}}
\author[2]{\fnm{Brian} \sur{Groenke}}
\author[2,3]{\fnm{Christof} \sur{Schötz}}
\author[4]{\fnm{Alessandro} \sur{Cotronei}}
\author[4]{\fnm{Eirik} \sur{Myrvoll-Nilsen}}
\author[2,3]{\fnm{Sebastian} \sur{Bathiany}}
\author[4]{\fnm{Martin} \sur{Rypdal}}
\author*[2,3,5]{\fnm{Niklas} \sur{Boers}}\email{n.boers@tum.de}

\affil[1]{\orgdiv{Department of Mathematics, School of Computation, Information and Technology}, \orgname{Technical University of Munich}, \orgaddress{\street{Boltzmannstraße 3}, \city{Garching}, \postcode{85748}, \state{Bavaria}, \country{Germany}}}

\affil[2]{\orgdiv{Research Domain IV -- Complexity Science}, \orgname{Potsdam Institute for Climate Impact Research}, \orgaddress{\street{Telegrafenberg A 31}, \city{Potsdam}, \postcode{14473}, \state{Brandenburg}, \country{Germany}}}

\affil[3]{\orgdiv{Munich Climate Center and Earth System Modelling Group, Department of Aerospace and Geodesy, TUM School of Engineering and Design}, \orgname{Technical University of Munich}, \orgaddress{\street{Lise-Meitner-Straße 9}, \city{Ottobrunn}, \postcode{85521}, \state{Bavaria}, \country{Germany}}}

\affil[4]{\orgdiv{Department of Mathematics and Statistics, Faculty of Science and Technology}, \orgname{UiT The Arctic University of Norway}, \orgaddress{\street{Hansine Hansens veg 18}, \city{Tromsø}, \postcode{9019}, \state{Troms}, \country{Norway}}}

\affil[5]{\orgdiv{Department of Mathematics and Statistics}, \orgname{University of Exeter}, \orgaddress{\street{North Park Road}, \city{Exeter}, \postcode{UK EX4 4QE}, \state{Devon}, \country{United Kingdom}}}

\maketitle

Ditlevsen and Ditlevsen \cite{Ditlevsen2023WarningCirculation} (DD23 hereafter) propose a statistical framework to estimate the timing of a potential collapse of the Atlantic Meridional Overturning Circulation (AMOC) based on extrapolating information from observed sea-surface temperature (SST) variability. By fitting a stochastic one-dimensional fold-bifurcation model to an SST-based fingerprint of the AMOC using Maximum Likelihood Estimation (MLE), they conclude that a collapse is most likely to occur in the middle of the 21st century, with a reported 95\% confidence interval covering the time span from 2037 to 2109. Given the profound implications of such a claim for both climate and society, it is essential to thoroughly test the robustness of this result, to critically assess the underlying assumptions and uncertainties, and to estimate the extent to which the reported confidence interval reflects the true limits of current knowledge.


Here we examine the sensitivity of DD23's results and argue that four types of uncertainty are insufficiently explored in their analysis: (i) structural uncertainty associated with the assumed low-order bifurcation model, (ii) statistical uncertainty in their model fit, (iii) uncertainty in the representativeness of SST-based fingerprints as proxies for the high-dimensional AMOC dynamics, and (iv) uncertainty in the underlying data, arising from non-stationary observational coverage and dataset preprocessing. 
Our results do not dispute the possibility that the AMOC might abruptly or irreversibly transition to an alternative state in response to anthropogenic forcing, nor that the stability of AMOC has already declined. However, using synthetic experiments and a systematic analysis of alternative fingerprints and observational products, we show that the tipping times estimated by DD23 are highly sensitive to the uncertainties listed above, and extend several millennia into the future when these uncertainties are thoroughly propagated. 

\section*{Modelling Assumptions and Structural Uncertainty}

The predictive framework introduced in DD23 is based on extrapolation of information from spatially averaged historical sea-surface temperature data, assuming an identical linear trend in forcing for past and future. It relies on a specific stochastic differential equation model:
\begin{align}
    \dint X_t &= -(A(X_t-m)^2+\lambda(t))\dint t + \sigma\,\dint B_t \label{eq: DD23} \\
    \lambda(t) &= \lambda_0(1-\Theta[t-t_0](t-t_0)/\tau_r), \nonumber
\end{align}
where $X_t$ represents the AMOC strength and $\lambda(t)$ denotes a linearly evolving external forcing that induces a tipping transition at time $t_c = t_0 + \tau_r$. DD23 devise a Maximum Likelihood Estimation (MLE) routine for the time of tipping $t_c$ and other model parameters. Although DD23 demonstrated that this approach provides a good statistical fit to their choice of SST-based AMOC fingerprint, its validity as a predictive tool hinges on the adequacy of the underlying physical assumptions. While we focus on the MLE routine, most points of criticism raised throughout this comment also apply to the running window approach alternatively used in DD23, which relies on the extrapolation of variance and autocorrelation trends.

The representation of the AMOC, a highly complex and spatially extended basin-scale ocean circulation system, as a one-dimensional stochastic equation with time-scale-independent noise, is problematic for predicting the time of a potential future AMOC tipping event. The mathematical justification provided by DD23 for using such a model as a conceptual representation of the AMOC is that all fold bifurcations are topologically equivalent to Eq.~\eqref{eq: DD23}. However, this is a local result valid only in an infinitesimal neighborhood of the bifurcation point \cite{Kuznetsov2004AppliedBifurcationTheory}. It may not account for AMOC dynamics during the historical time period, which can deviate significantly from the normal form.

\begin{figure}[t]
\centering
\includegraphics[width=\textwidth]{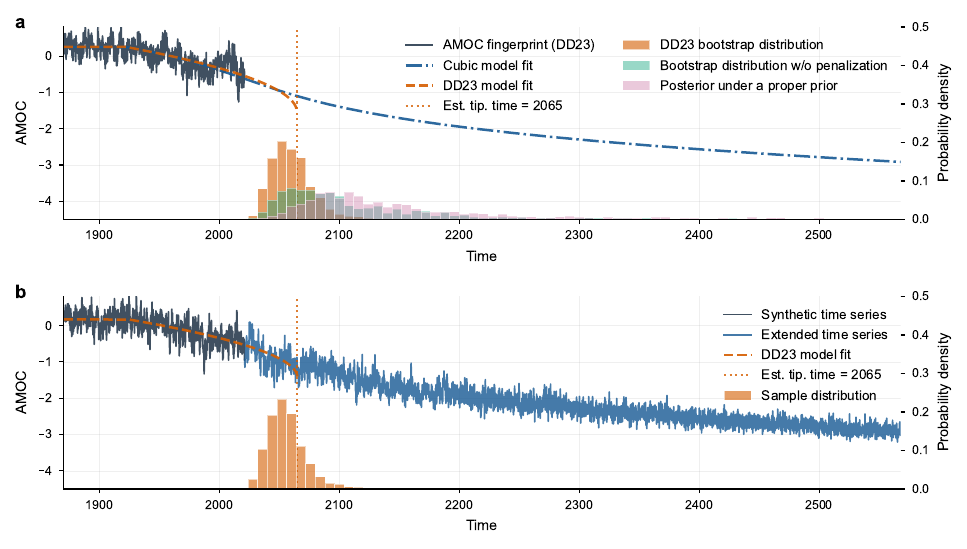}
\caption{\textbf{Structural uncertainty in tipping-time estimation.} (a) Analysis of the AMOC fingerprint proposed by DD23 using both their MLE method and an adaptation with a third-order drift term. If the bifurcation model is chosen such that tipping is not mechanically prescribed but optional (switching from a second-order to a third-order drift term, see Eqs.~\eqref{eq: quad} and \eqref{eq: cubic} in the Methods below), the MLE fitting routine suggests no tipping in finite time at all. In contrast to the originally proposed drift by DD23, the third-order drift term has the additional advantage that it in principle allows for two alternative AMOC stable states. Distributions are obtained by (i) the bootstrap routine proposed by DD23 for uncertainty quantification, (ii) the same bootstrap routine without using a penalization on the parameter $A$ (an accumulation of estimates around the year 2800 is not depicted here), and (iii) a Bayesian MCMC routine using a statistically proper and less informative prior on $A$.
(b) Synthetic AMOC time series generated from the model with third-order drift inferred in (a). Applying the DD23 method, assuming the second-order drift (Eq.~\eqref{eq: quad}), to the black portion of these time series produces spurious tipping-time estimates around the mid-21st century, despite the absence of a true transition. The histogram shows the sample distribution of this estimation over 1000 independent realizations of the model time series.}\label{fig: ModelUncertainty}
\end{figure}

To probe how strongly the DD23 inference depends on the assumed structure of the model, we extend their maximum-likelihood framework by replacing the quadratic fold normal form with a more general cubic approximation of the dynamics (see Methods and Cotronei et al.~\cite{Cotronei2026AMOCTippingUncertainty} for details on this adaptation). 
A cubic drift, i.e.~a third-order deterministic term in the model equation \eqref{eq: DD23}, is the minimal polynomial extension that can represent both a true fold bifurcation and a reversible, smooth shift of the stable state, without imposing the existence of a tipping point a priori as in the model of DD23. 
The model structure with a cubic drift also arises naturally from the reduced Stommel-Cessi box model of the thermohaline circulation \cite{Cessi1994BoxModel} and, importantly, permits multiple stable states when a tipping point is present. The model with quadratic drift, as proposed in DD23, on the other hand, necessarily has a bifurcation point for all parameter choices and permits only one stable state.
When the alternative third-order model is used to construct an MLE for the slope of the ramping parameter $\lambda$, in the same vein as DD23, the observed AMOC fingerprint is fitted equally well, if not better (measured by residual squared errors, a visual inspection of the residual quantile-quantile plot, and model comparisons via AIC and BIC, which favor the cubic model with a difference of 10 and 5 in the two criteria, respectively). However, the extrapolated future trajectory changes qualitatively: instead of an imminent fold-type collapse, the best-fit cubic model yields a smooth continuation of the weakening trend (see Fig.~\ref{fig: ModelUncertainty}a). 
The divergence between the quadratic and cubic projections reflects structural model uncertainty; different, equally plausible low-order representations of the drift lead to fundamentally different assessments of future tipping risks. 
This demonstrates that the DD23 tipping estimate relies on imposing a normal-form fold-bifurcation geometry a priori, rather than that geometry being uniquely determined from the data.

The consequences of such a structural mismatch are further illustrated in Fig.~\ref{fig: ModelUncertainty}b. 
Here we apply the original DD23 MLE framework to synthetic time series generated from the best-fit cubic model inferred in Fig.~\ref{fig: ModelUncertainty}a (see Methods for details). 
By construction, these realizations approximately reproduce the statistical properties of the observed AMOC fingerprint (such as the evolution of its mean state, variance, and autocorrelation, with 75\% of realizations passing the significance test of DD23 in terms of their increases above baseline) but do not exhibit a true tipping point when evolved forward in time.
Nevertheless, when the DD23 fold-bifurcation model is fitted to the observed period of each realization of the model with third-order drift, the likelihood optimization systematically yields finite tipping times clustered around the mid-21st century. This behavior demonstrates that the DD23 framework can produce spurious tipping-time estimates when applied to data generated by a model not exhibiting a bifurcation in the present parameter configuration, and hence not exhibiting tipping dynamics. While the model with third-order drift deviates from the normal-form structure assumed by DD23, its sample time series still pass goodness-of-fit metrics with respect to the normal-form structure, like the residual quantile-quantile analysis proposed by DD23. 

In addition to the specific points regarding structural uncertainty discussed above, several other possible model deviations should be discussed. For example, alternative choices of the stochastic term in Eq.~\eqref{eq: DD23} can lead to non-stationary time-series statistics independently of any AMOC weakening \cite{Morr2024RedNoiseCSD, Morr2024KramersMoyalEWS}. We point the reader to Figure 1 in Ben-Yami et al.~\cite{Ben-Yami2024TippingTime}, which shows that DD23's method predicts a tipping time even for a linear system subjected to non-linear forcing and non-stationary correlated noise.
It is also plausible that the recent downward curvature in the AMOC SST fingerprint time series used by DD23 just reflects the impact of a strongly non-linear external forcing, for example, the global warming trend that has accelerated during the 20th century, or non-linear increases in freshwater inflow to the North Atlantic. This trend in forcing may not extrapolate well into the future, thus biasing estimates of the tipping time under their model. 

\section*{Failures of the Confidence Assessment}

One of the main results of DD23 is the prediction that AMOC tipping will occur ``with high confidence ... as soon as mid-century'' \citep[p.7]{Ditlevsen2023WarningCirculation}, supported by a 95\% confidence interval (CI) for the tipping time of 2037--2109 (revised from 2025--2095 in a correction). We identify two reasons why this statement of confidence is misleading: First, we show via a simulation study that their CI procedure, based on a penalized MLE and parametric bootstrap, does not achieve the nominal 95\% uniform coverage across the full parameter space, in fact yielding 0\% coverage under the original implementation and 45\% under our improved implementation for some parameter configurations. Second, we argue that, even if nominal coverage were to be achieved, DD23's interpretation of their CI as providing a likely range of future tipping times (see Figure 7 caption in \cite{Ditlevsen2023WarningCirculation}) is not justified under their chosen statistical framework. All numerical experiments in this section build on the original stochastic differential equation model of DD23 in Eq.~\eqref{eq: DD23}

Regarding the first reason, we note that a 95\% CI procedure for the tipping time parameter $t_c$ is defined for a family of data generating processes $\mathbb{P}_{\phi,t_c}$ indexed by unknown parameters $(\phi,t_c) \in \Phi$ as any method of computing two numbers $L$ and $R$ from the generated data such that the probability of $L$ being smaller and $R$ being larger than $t_c$ is at least 95\% uniformly over all parameter vectors $(\phi,t_c) \in \Phi$. This can be stated formally as
\begin{equation*}
    \inf_{(\phi,t_c)\in\Phi}\mathbb{P}_{\phi,t_c}(L \leq t_c \leq R) \geq 95\%\,.
\end{equation*}
In DD23, $\phi = (A, m, \lambda_0, \sigma, t_0)$ and $\mathbb{P}_{\phi,t_c}$ is defined by Equation \eqref{eq: DD23} using $\tau_r = t_c - t_0$. No non-trivial restrictions are given for the parameter space $\Phi$. The CI procedure $(L,R)$ is based on a parametric bootstrap using an MLE with a data-driven penalty on the parameter $A$.

\begin{figure}
    \centering
    \includegraphics[width=\linewidth]{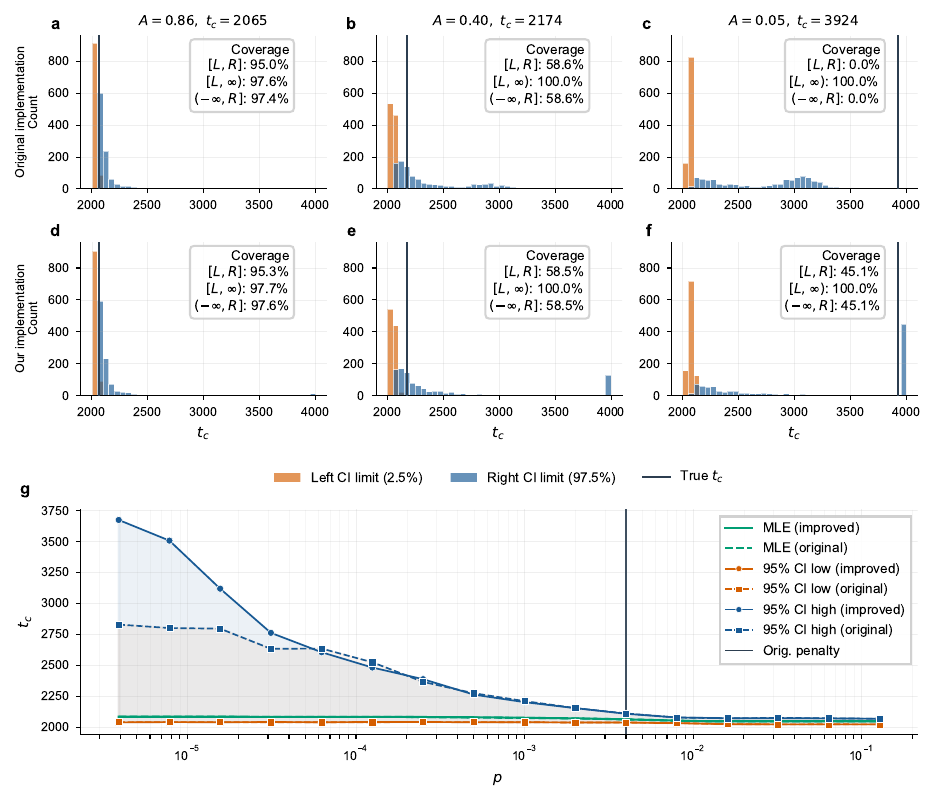}
    \caption{Evaluation of the confidence interval procedure of DD23. (a--f) Each panel shows histograms of the lower (orange) and upper (blue) confidence limits across 1000 simulation repetitions calculated using the original implementation (top row) or our implementation (bottom row). The panel columns correspond to different ground truth parameter vectors of the simulation, with the true critical time $t_c$ marked by a vertical black line. All values $>4000$ are collected in the rightmost bin of each panel. The text box in each panel reports the empirical coverage for the corresponding parameter vector, broken down into the two-sided centered interval and the two one-sided intervals. (g) MLE and bootstrap confidence intervals as a function of penalty weight. We compute the penalized MLE (green) for the critical time $t_c$ and the parametric bootstrap 95\% confidence intervals (orange and blue) from DD23 on the observed time series across a range of penalty weights $p$ using the original implementation and our improved one. Both orange and green lines coincide in this plot, respectively. The penalty weight $p = 0.004$ adopted in DD23 is indicated in black. The upper confidence limit exhibits strong sensitivity to the choice of penalty weight.}
    \label{fig:coverageSimulation}
\end{figure}

Using Monte-Carlo integration, we exactly follow this bootstrapping procedure and approximate $\mathbb{P}_{\phi,t_c}(L \leq t_c \leq R)$ for selected parameter vectors $(\phi,t_c)$, see Figure \ref{fig:coverageSimulation} and Methods. Specifically, we choose $m, \lambda_0, \sigma, t_0$ to fit the observational time series in DD23 and vary $A$ and $t_c$. For $t_c = 3924$ (equivalently $\tau_r = 2000$) and $A = 0.05$, we obtain $\mathbb{P}_{\phi,t_c}(L \leq t_c \leq R) \approx 0$ using the original code for the CI procedure. Since the original implementation directly imposes $A \geq 0.1$ through an ad hoc constraint, we additionally run a modified implementation of the CI procedure that still enforces the requirement $A>0$ while allowing values below $0.1$ (see Methods) and obtain $\mathbb{P}_{\phi,t_c}(L \leq t_c \leq R) \approx 45\%$. Hence, the method used in DD23 is not a 95\% CI procedure and can severely underestimate sampling uncertainty. Moreover, for most parameter configurations considered, the left limit $L$ is lower than the true parameter $t_c$ with probability $\approx 100\%$. Thus, any non-conservative CI procedure $(L^*, R^*)$ for a centered 95\% confidence interval, $\mathbb{P}_{\phi,t_c}(L^* \leq t_c) = \mathbb{P}_{\phi,t_c}(t_c \leq R^*) = 97.5\%$, would necessarily yield larger lower (and upper) bounds for these parameters.

While the apparent bias of the CI procedure depends on the parameters $A$ and $t_c$ and appears to diminish when evaluated at the parameter estimates reported in DD23 ($A=0.86$; $t_c = 2065$, equivalently $\tau_r = 141$), the failure of the CI procedure is not confined to extreme cases (Figure \ref{fig:coverageSimulation}b,c,e,f). For example, with $A=0.4$ and $t_c = 2174$ (equivalently $\tau_r = 250$) the procedure yields intervals whose endpoints are at most as large as the estimated ones, $2037$ and $2109$, respectively, with $\approx15\%$ probability. This occurs despite the true value being 2174, in clear violation of the 95\% confidence requirement. We therefore find no principled restriction of the parameter space for the observed data under which the procedure achieves valid coverage. Applying the parametric bootstrap with the unpenalized MLE produces coverage much closer to the nominal 95\% in our simulations. In the considerations further below and in the Methods section, we show that this penalization introduces an ad hoc but highly influential hyperparameter (Fig.~\ref{fig:coverageSimulation}g). For the observational time series used by DD23, the unpenalized MLE 95\% CI for $t_c$ is 2042--2834 using the original implementation (see green distribution in Fig.~\ref{fig: ModelUncertainty}) and 2042--5108 using our improved implementation. These CIs should be regarded as more accurately capturing the sampling uncertainty of the MLE method.

Regarding the second reason, we argue that, even if the CI were to achieve proper coverage, DD23's ``high confidence'' predictions that tipping will occur within the reported interval would still be misleading. The 95\% coverage probability of a CI only describes the long-run performance of the procedure used to generate the CI under repeated sampling from the data generating process. This is generally not the same as estimating the probability (or plausibility) that the true value of the parameter (here tipping time) is contained within an interval computed from a single sample \citep{neymanOutlineTheoryStatistical1937,moreyFallacyPlacingConfidence2015}. Therefore, colloquial claims of ``confidence'', ``likelihood'', or otherwise about the future tipping time based on a specific CI, such as DD23's statement that ``transition of the AMOC is most likely to occur around 2037--2109 (95\% confidence interval)'' \citep[][p.2]{Ditlevsen2023WarningCirculation}, are not supported by the underlying statistical theory. 

This could be partially addressed by instead adopting a Bayesian interpretation of the model and estimating the posterior distribution over the tipping time, given the observed fingerprint. However, this would in turn require formal specification of a prior distribution over all parameters of the model, including the tipping time itself. While confidence intervals generated by the parametric bootstrap may, under certain conditions, approximate the Bayesian posterior \citep{efronBayesianInferenceParametric2012}, this correspondence cannot generally be assumed without mathematical or empirical verification. Furthermore, under the Bayesian interpretation, a robust probabilistic forecast of future tipping time would need to include a sensitivity analysis of the predictions to both the choice of model and prior \citep{bergerOverviewRobustBayesian1994}.

This is here particularly relevant since the aforementioned penalty term in the likelihood effectively acts as an improper but highly informative prior on the parameter $A$ in Eq. \eqref{eq: DD23}.
We show that a Bayesian interpretation confirms that DD23's inferences are sensitive to this choice by using Markov Chain Monte Carlo to obtain an estimate of the posterior distribution over tipping times under an alternative choice of prior for $A$ (see Methods). The alternative prior fulfills the same basic function as DD23's penalty by down-weighting small values of $A$ while being statistically proper and only weakly informative. We use noninformative uniform priors for all other parameters, including the tipping time.
The resulting 95\% credible range of tipping times is then 2055--2512 with median $t_c = 2122$ (Fig.~\ref{fig: ModelUncertainty}). This demonstrates that, even under a generous Bayesian interpretation, DD23 still considerably underestimate statistical uncertainty within their own model.

\section*{Representativeness of Proxies for Underlying Dynamics}

\begin{figure}[t]
\centering
\includegraphics[width=\textwidth]{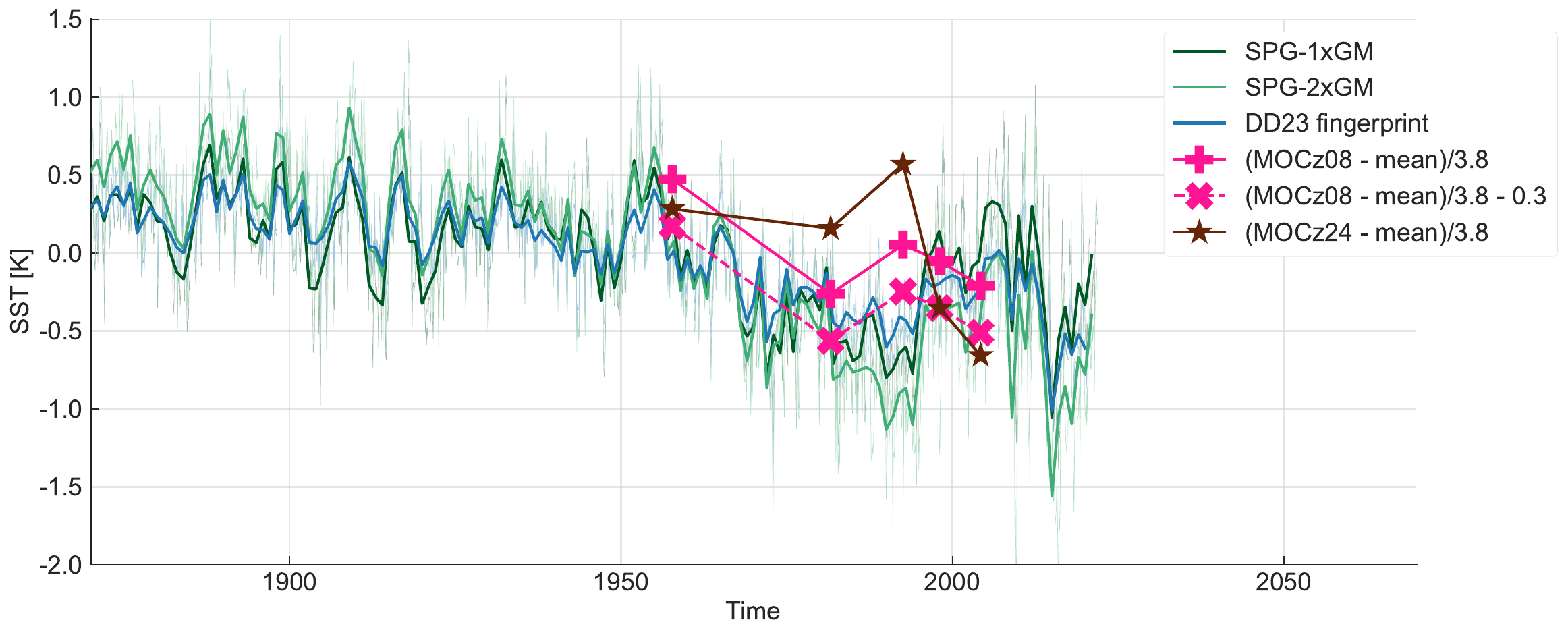}
\caption{\textbf{Comparison of SST fingerprints and MOC$_z$ data} 
The MOC$_z$ data \cite{bryden2005Slowing} comprises five values for the AMOC streamfunction at 26$^\circ$N calculated from hydrographic sections. Kanzow et al. 2010\cite{Kanzow2010Seasonal} corrected the seasonal aliasing of the observations using RAPID data from 2004-2008 (MOC$_z$08), which DD23 compare to the SST fingerprints by converting to K using a 3.8Sv/K factor \cite{Caesar2018AMOCFingerprint} (pink pluses, solid line). To visually reproduce DD23's Figure 7, we shifted the data down by approximately 0.3K (pink exes, dotted line). We additionally correct the seasonal aliasing with RAPID data from 2004-2024 (MOC$_z$24, brown star, solid line). While DD23 report that the best fit to the shifted MOC$_z$08 data is approximately their SPG-2xGMT fingerprint (blue), when the correct data processing method (see Methods) is used, the best fit is approximately SPG-1xGMT (dark green), not SPG-2xGMT (light green). The annual mean (thick line) is plotted over the monthly fingerprint values (thin line).
}\label{fig: DD_fingerprint}
\end{figure}

DD23 rely on a specific sea-surface temperature (SST) fingerprint, based on averaging SSTs over the subpolar gyre (SPG) region. We note that in calculating the fingerprint they omitted a step in the latitudinal area weighting of the gridded SST data, and included grid cells with sea ice coverage where SSTs are at the freezing temperature and hence not informative of ocean heat transport (see Methods and Figure \ref{fig: DD_fingerprint}). When these issues are corrected, DD23's fingerprint does not pass their initial test for significantly high variance and lag-one autocorrelation. More importantly, they subtract $2\times$ the global mean SST to obtain their specific one-dimensional proxy for the Atlantic Meridional Overturning Circulation (AMOC) strength. This differs from the established version of the SPG AMOC fingerprint introduced by \cite{Caesar2018AMOCFingerprint}, where only $1\times$ the global mean is subtracted. DD23 justify this different fingerprint with two arguments: First, they contend that due to polar amplification, the warming trend in the SPG region is twice that of the global mean temperature increase, and thus more must be subtracted. Second, DD23 optimize $\text{SST~}_\text{SPG}-\beta ~\text{SST}_\text{global}$ with respect to $\beta$ by fitting the fingerprint to hydrographic section data, and quoting $\beta=1.95\approx2$ as the optimal value for the coefficient. 

The physical basis of the AMOC SPG SST fingerprint is that a reduction in AMOC strength causes the SPG SSTs to cool relative to how much they would have warmed without AMOC weakening. Thus, the subtraction of some coefficient of the global mean SST timeseries needs to account for the amplification of the global SST warming trend at the SPG's latitudes. To calculate a rough approximation of this coefficient, we may consider the warming at the same latitude band of SSTs in the Pacific Ocean relative to the global SST warming trend. When we average the SST trend amplification for the SST dataset used by DD23 (1870-2021 HadISST) in the area 40-65$^\circ$N and 130$^\circ$E-110$^\circ$W, the value is $\beta=1.02$, much smaller than the twofold amplification used by DD23 to justify the $2\times$ global mean subtraction. 

Regarding the justification of a factor 2 via more direct observations, DD23's Figure 7 \cite{Ditlevsen2023WarningCirculation} uses the seasonally corrected hydrographic sections from Kanzow et al. 2010 \cite{Kanzow2010Seasonal} (Figure \ref{fig: DD_fingerprint}, solid pink), which DD23 refer to as the "MOC$_z$" data (citing \cite{Frajka-Williams2019AtlanticVariability}). It should be noted that there exists no obvious way to convert monthly AMOC streamfunction data to an SST value. DD23 use the 3.8 Sv/K
slope from Figure 5 of \cite{Caesar2018AMOCFingerprint}, which was introduced as a conversion rate between centennial AMOC streamfunction trends to centennial SST trends. Furthermore, Figure \ref{fig: DD_fingerprint} shows two more possible versions of the MOC$_z$ data. The first (dashed pink) is the same converted data shifted down by 0.3 K, to make it visually similar to the data plotted in DD23's Figure 7. The second (solid purple) is an updated seasonal correction of the hydrographic section observations, using 2004-2024 RAPID-MOCHA data \cite{RAPID} instead of the 2004–2008 data used by \cite{Kanzow2010Seasonal}. None of these versions of the MOC$_z$ data are a good fit to any version of the SST fingerprint (Figure \ref{fig: DD_fingerprint}). In particular, when we minimize $\text{SST~}_\text{SPG}-\beta~ \text{SST}_\text{global}$ with respect to $\beta$ using our own $\text{SST~}_\text{SPG}$ and $\text{SST}_\text{global}$ calculations, the best fit for the centred MOC$_z$ data from \cite{Kanzow2010Seasonal} is $\beta=0.29$, and the best fit for the 0.3K shifted data is $\beta=0.92$ (Figure \ref{fig: DD_fingerprint}). None of these calculations should be taken to mean that subtracting one time the global mean SST from the SPG SSTs results in an optimal AMOC fingerprint. They demonstrate, however, that we find no compelling reason to modify the traditional SPG AMOC fingerprint by subtracting a different coefficient of the global mean SST. 

More generally, the AMOC is a complex, spatially extended system with many degrees of freedom, and it is an oversimplification to assume its future tipping time is fully encoded in a single one-dimensional temperature time series. 
While the ``warming hole" in the North Atlantic has been linked to AMOC weakening \cite{Rahmstorf2015ExceptionalCirculation, Caesar2018ObservedCirculation, Drijfhout2012IsPatterns, Menary2018AnModels, Liu2020ClimateClimate}, the correlation between the SPG-based index and the actual AMOC streamfunction is non-stationary and sensitive to the time period and forcing scenario considered \cite{Little2020DoStrength, Jackson2018HysteresisGCM}. This non-stationarity is plausible because the North Atlantic warming hole is not solely caused by ocean circulation; it likely results from a complex interplay of ocean heat transport, atmospheric forcing, and aerosol-induced radiative anomalies \cite{Li2022Century-longExplanation, He2022ACirculation, Ferster2022SlowdownDecline, Ghosh2022TwoWarming, Keil2020MultipleHole}. This is further complicated by recent observations from the OSNAP program, which suggest that the SPG region plays a different role in deep water formation than previously assumed \cite{Lozier2019AAtlantic, Chafik2022IrmingerVariability}, casting doubt on the reliability of this fingerprint as a predictive tool. Additionally, evidence suggests that the SPG itself may possess an independent tipping point separate from the AMOC \cite{Sgubin2017AbruptModels, Swingedouw2021OnModels, McKay2022ExceedingPoints}. Consequently, as DD23 themselves concede, the early warning signals (EWS) detected in the SPG-based index may represent a localized transition rather than a collapse of the global overturning circulation \cite{BenYami2023AMOCDataCSD}.

\begin{figure}[t]
\centering
\includegraphics[width=\textwidth]{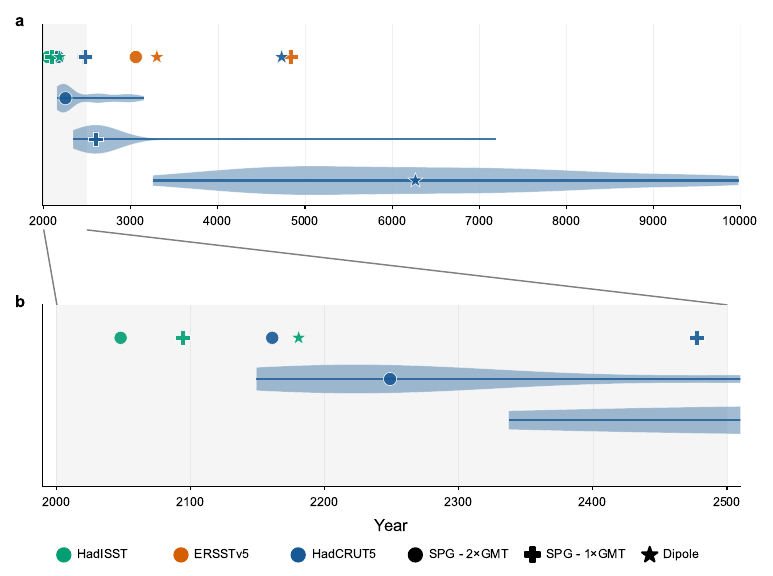}
\caption{\textbf{Range of tipping times in alternative fingerprints and data.} Tipping times estimated using DD23's MLE method. The best estimate of the tipping time is calculated for the classical SPG-based AMOC index (circle), the fingerprint used by DD23 (plus), and the Dipole index (star). We use three different observational SST datasets for this analysis: HadISST1 (turquoise), ERSSTv5 (orange), and HadCRUT5 (blue). Additionally, the blue violins display the tipping times for each of the 200-member uncertainty ensembles of HadCRUT5-derived fingerprints. Note that the modified fingerprint by DD23 consistently shows the earliest tipping time, partly due to subtracting $2\times$ the global mean SST rather than only one time. This Figure is similar to Fig.~4 in Ben-Yami et al.~\cite{Ben-Yami2024TippingTime}, accounting for the error-corrected version of DD23's MLE method and now also determining the optimal penalization term for each individual time series.
}\label{fig: DataUncertainties}
\end{figure}

When the MLE method of DD23 is applied to the dipole index \cite{Roberts2013ACirculation, Jackson2020FingerprintsAMOC}, an alternative physically motivated AMOC fingerprint, the resulting tipping-time estimates span centuries to millennia (see Fig.~\ref{fig: DataUncertainties}). When calculated from the same SST dataset as DD23's data, the conventionally used fingerprint constructed by subtracting $1\times$ rather than $2\times$ the global mean SST from the SPG SST \cite{Caesar2018AMOCFingerprint} yields an estimated tipping time around 2100, as noted by DD23 themselves. We emphasize that many other fingerprints could be constructed \cite{Jackson2020FingerprintsAMOC}, and no single choice is uniquely justified.

\section*{Dataset Preprocessing and Non-stationary Coverage}

The reliability of tipping time prediction is further undermined by potential biases in the historical climate data used. In particular, methods for predicting tipping times, including the MLE approach used by DD23, build on higher-order statistics like variance and autocorrelation, which are often unreliable due to the non-stationarity of observational coverage and the preprocessing methods used to fill data gaps \cite{Smith2023ReliabilitySeries, Boers2023ReplyCirculation, BenYami2023AMOCDataCSD}.

Specifically, historical SST records have seen an exponential growth in the number of observations \cite{Lundstad2023TheHCLIM, Kennedy2019AnSet}. In earlier years, measurements were sparse and concentrated in specific areas, requiring sophisticated infilling procedures to produce globally complete datasets. The HadISST1 dataset used by DD23 employs Reduced Space Optimal Interpolation (RSOI) \cite{Rayner2003GlobalCentury}, which regularizes the fit toward the mean in regions and times where less data is available. This process has the effect of damping the variance in the earlier part of the record where observations are more sparse. As observational density increases toward the mid-20th century, the captured variability grows, resulting in an artificial trend in variance (and potentially other higher-order statistics) that can be misinterpreted as a signal of critical slowing down and, hence, of stability loss \cite{Rayner2003GlobalCentury, BenYami2023AMOCDataCSD, Smith2023ReliabilityResilienceEstimation}.

The sensitivity of tipping time estimates to these artifacts becomes clear when comparing different observational products. Alternative datasets, such as ERSSTv5 \cite{Huang2017ExtendedIntercomparisons} and HadCRUT5 \cite{Morice2021AnSet}, employ different infilling and bias-correction procedures, leading to different trends in variance and autocorrelation \cite{Ben-Yami2024TippingTime}. Applying the DD23 methodology to fingerprints constructed from the ERSSTv5 or HadCRUT5 datasets yields a range of tipping times spanning thousands of years (see Fig.~\ref{fig: DataUncertainties}). Given that the variance increase in HadISST1 is known to be at least partially artificial, any prediction relying on the precise evolution of this statistic lacks the necessary robustness to support a definitive warning of collapse in the mid-21st century.

Furthermore, applying the tipping-time estimation to the HadCRUT5 uncertainty ensemble yields a similarly wide spread of inferred collapse times (violins in Fig.~\ref{fig: DataUncertainties}). Because this ensemble is designed to reflect fundamental limitations in historical observational accuracy, the resulting dispersion in tipping-time estimates represents an irreducible source of uncertainty. Such uncertainty cannot be curtailed by improved statistical fitting alone and should therefore be explicitly incorporated into any comprehensive uncertainty assessment.

\section*{Concluding Remarks}

The results presented here highlight that the tipping-time estimates reported in DD23 strongly depend on specific modelling assumptions and data choices. The uncertainty ranges associated with these assumptions and choices are only marginally assessed by DD23, leading to reported confidence intervals in their study that are considerably biased and too narrow. When structural uncertainty in the dynamical model, parametric and statistical uncertainties, ambiguity in the physical meaning of SST-based AMOC fingerprints, and non-stationary observational artifacts and measurement uncertainties are taken into account, the inferred collapse times span several millennia. In addition, we have shown that using an equally plausible model leads to a configuration without tipping, for which DD23's method would still falsely predict a tipping time in the near future. The narrow confidence intervals reported in DD23 reflect uncertainty conditional on all the choices discussed above, rather than the full uncertainty relevant for real-world prediction.

Our analysis does not preclude the possibility of a future AMOC collapse in any way, nor does it diminish the importance of continued monitoring of AMOC stability and immediate action to mitigate the risks posed by anthropogenic climate change. However, it demonstrates that current observational records and simplified statistical models do not support a meaningful quantitative prediction of a tipping time. Claims of a specific time frame of collapse should therefore not be communicated to the public or policymakers, who could mistake them for reliable or actionable knowledge.

\vspace{1em}
\noindent\textbf{Competing Interests:} The authors declare no competing interests.

\vspace{1em}
\noindent\textbf{Data Availability:} All code for the discussed data analyses and simulation studies are available upon request.

\vspace{1em}
\noindent\textbf{Author Contributions:} AM, MBY, BG, AC, CS, EMN, SB, MR, and NB conceived the study. AM, MBY, BG, AC, CS, and EMN developed the methods. AM, MBY, BG, and CS carried out the analyses. All authors contributed to the writing and revision of the manuscript.

\vspace{1em}
\noindent\textbf{Acknowledgements:}
This is ClimTip contribution \#146; the ClimTip project has received funding from the European Union's Horizon Europe research and innovation programme under grant agreement No. 101137601. This project is funded by the Advanced Research + Invention Agency (ARIA). NB acknowledges additional funding by the VolkswagenFoundation. B.G. acknowledges funding from the Past to Future (P2F) project under the European Union’s Horizon Europe research and innovation programme grant agreement No. 101184070: views and opinions are solely that of the authors and do not necessarily reflect those of the European Union or the European Climate, Infrastructure and Environment Executive Agency (CINEA). The authors gratefully acknowledge the Ministry of Research, Science and Culture (MWFK) of Land Brandenburg for supporting this project by providing resources on the high performance computer system at the Potsdam Institute for Climate Impact Research.

\bibliography{BibAll, Ben-YamiBib, Groenke.bib}


\begin{thebibliography}{48}
\ifx \bisbn   \undefined \def \bisbn  #1{ISBN #1}\fi
\ifx \binits  \undefined \def \binits#1{#1}\fi
\ifx \bauthor  \undefined \def \bauthor#1{#1}\fi
\ifx \batitle  \undefined \def \batitle#1{#1}\fi
\ifx \bjtitle  \undefined \def \bjtitle#1{#1}\fi
\ifx \bvolume  \undefined \def \bvolume#1{\textbf{#1}}\fi
\ifx \byear  \undefined \def \byear#1{#1}\fi
\ifx \bissue  \undefined \def \bissue#1{#1}\fi
\ifx \bfpage  \undefined \def \bfpage#1{#1}\fi
\ifx \blpage  \undefined \def \blpage #1{#1}\fi
\ifx \burl  \undefined \def \burl#1{\textsf{#1}}\fi
\ifx \doiurl  \undefined \def \doiurl#1{\url{https://doi.org/#1}}\fi
\ifx \betal  \undefined \def \betal{\textit{et al.}}\fi
\ifx \binstitute  \undefined \def \binstitute#1{#1}\fi
\ifx \binstitutionaled  \undefined \def \binstitutionaled#1{#1}\fi
\ifx \bctitle  \undefined \def \bctitle#1{#1}\fi
\ifx \beditor  \undefined \def \beditor#1{#1}\fi
\ifx \bpublisher  \undefined \def \bpublisher#1{#1}\fi
\ifx \bbtitle  \undefined \def \bbtitle#1{#1}\fi
\ifx \bedition  \undefined \def \bedition#1{#1}\fi
\ifx \bseriesno  \undefined \def \bseriesno#1{#1}\fi
\ifx \blocation  \undefined \def \blocation#1{#1}\fi
\ifx \bsertitle  \undefined \def \bsertitle#1{#1}\fi
\ifx \bsnm \undefined \def \bsnm#1{#1}\fi
\ifx \bsuffix \undefined \def \bsuffix#1{#1}\fi
\ifx \bparticle \undefined \def \bparticle#1{#1}\fi
\ifx \barticle \undefined \def \barticle#1{#1}\fi
\bibcommenthead
\ifx \bconfdate \undefined \def \bconfdate #1{#1}\fi
\ifx \botherref \undefined \def \botherref #1{#1}\fi
\ifx \url \undefined \def \url#1{\textsf{#1}}\fi
\ifx \bchapter \undefined \def \bchapter#1{#1}\fi
\ifx \bbook \undefined \def \bbook#1{#1}\fi
\ifx \bcomment \undefined \def \bcomment#1{#1}\fi
\ifx \oauthor \undefined \def \oauthor#1{#1}\fi
\ifx \citeauthoryear \undefined \def \citeauthoryear#1{#1}\fi
\ifx \endbibitem  \undefined \def \endbibitem {}\fi
\ifx \bconflocation  \undefined \def \bconflocation#1{#1}\fi
\ifx \arxivurl  \undefined \def \arxivurl#1{\textsf{#1}}\fi
\csname PreBibitemsHook\endcsname

\bibitem[\protect\citeauthoryear{Ditlevsen and
  Ditlevsen}{2023}]{Ditlevsen2023WarningCirculation}
\begin{barticle}
\bauthor{\bsnm{Ditlevsen}, \binits{P.D.}},
\bauthor{\bsnm{Ditlevsen}, \binits{S.}}:
\batitle{{Warning of a forthcoming collapse of the Atlantic meridional
  overturning circulation}}.
\bjtitle{Nature Communications}
\bvolume{14}(\bissue{1}),
\bfpage{4254}
(\byear{2023})
\doiurl{10.1038/s41467-023-39810-w}
\end{barticle}
\endbibitem

\bibitem[\protect\citeauthoryear{Kuznetsov}{2004}]{Kuznetsov2004AppliedBifurcationTheory}
\begin{bbook}
\bauthor{\bsnm{Kuznetsov}, \binits{Y.A.}}:
\bbtitle{Elements of {Applied} {Bifurcation} {Theory}},
\bedition{4}th edn.
\bsertitle{Applied {Mathematical} {Sciences}},
vol. \bseriesno{112}.
\bpublisher{Springer},
\blocation{New York}
(\byear{2004}).
\doiurl{10.1007/978-1-4757-3978-7}
\end{bbook}
\endbibitem

\bibitem[\protect\citeauthoryear{Cotronei
  et~al.}{2026}]{Cotronei2026AMOCTippingUncertainty}
\begin{barticle}
\bauthor{\bsnm{Cotronei}, \binits{A.}},
\bauthor{\bsnm{{Myrvoll-Nilsen}}, \binits{E.}},
\bauthor{\bsnm{Rypdal}, \binits{M.}}:
\batitle{Evaluating model uncertainty in critical threshold estimations from
  time series data: Application to the {{Atlantic}} meridional {{Overturning
  Circulation}}}.
\bjtitle{Frontiers in Climate}
\bvolume{8},
\bfpage{1761461}
(\byear{2026})
\doiurl{10.3389/fclim.2026.1761461}
\end{barticle}
\endbibitem

\bibitem[\protect\citeauthoryear{Cessi}{1994}]{Cessi1994BoxModel}
\begin{barticle}
\bauthor{\bsnm{Cessi}, \binits{P.}}:
\batitle{A simple box model of stochastically forced thermohaline flow}.
\bjtitle{Journal of Physical Oceanography}
\bvolume{24}(\bissue{9}),
\bfpage{1911}--\blpage{1920}
(\byear{1994})
\doiurl{10.1175/1520-0485(1994)024<1911:ASBMOS>2.0.CO;2}
\end{barticle}
\endbibitem

\bibitem[\protect\citeauthoryear{Morr and Boers}{2024}]{Morr2024RedNoiseCSD}
\begin{botherref}
\oauthor{\bsnm{Morr}, \binits{A.}},
\oauthor{\bsnm{Boers}, \binits{N.}}:
Detection of {Approaching} {Critical} {Transitions} in {Natural} {Systems}
  {Driven} by {Red} {Noise}.
Physical Review X
\textbf{14}(2)
(2024)
\doiurl{10.1103/PhysRevX.14.021037}
\end{botherref}
\endbibitem

\bibitem[\protect\citeauthoryear{Morr et~al.}{2024}]{Morr2024KramersMoyalEWS}
\begin{barticle}
\bauthor{\bsnm{Morr}, \binits{A.}},
\bauthor{\bsnm{Riechers}, \binits{K.}},
\bauthor{\bsnm{Gorjão}, \binits{L.R.}},
\bauthor{\bsnm{Boers}, \binits{N.}}:
\batitle{Anticipating critical transitions in multidimensional systems driven
  by time- and state-dependent noise}.
\bjtitle{Physical Review Research}
\bvolume{6}(\bissue{3}),
\bfpage{033251}
(\byear{2024})
\doiurl{10.1103/PhysRevResearch.6.033251}
\end{barticle}
\endbibitem

\bibitem[\protect\citeauthoryear{Ben-Yami
  et~al.}{2024}]{Ben-Yami2024TippingTime}
\begin{barticle}
\bauthor{\bsnm{Ben-Yami}, \binits{M.}},
\bauthor{\bsnm{Morr}, \binits{A.}},
\bauthor{\bsnm{Bathiany}, \binits{S.}},
\bauthor{\bsnm{Boers}, \binits{N.}}:
\batitle{Uncertainties too large to predict tipping times of major {Earth}
  system components from historical data}.
\bjtitle{Science Advances}
\bvolume{10}(\bissue{31}),
\bfpage{4841}
(\byear{2024})
\doiurl{10.1126/sciadv.adl4841}
\end{barticle}
\endbibitem

\bibitem[\protect\citeauthoryear{Neyman}{1937}]{neymanOutlineTheoryStatistical1937}
\begin{barticle}
\bauthor{\bsnm{Neyman}, \binits{J.}}:
\batitle{Outline of a {{Theory}} of {{Statistical Estimation Based}} on the
  {{Classical Theory}} of {{Probability}}}.
\bjtitle{Philosophical Transactions of the Royal Society of London, Series A:
  Mathematical and Physical Sciences}
\bvolume{236}(\bissue{767}),
\bfpage{333}--\blpage{380}
(\byear{1937})
\doiurl{10.1098/rsta.1937.0005}
\end{barticle}
\endbibitem

\bibitem[\protect\citeauthoryear{Morey
  et~al.}{2015}]{moreyFallacyPlacingConfidence2015}
\begin{barticle}
\bauthor{\bsnm{Morey}, \binits{R.D.}},
\bauthor{\bsnm{Hoekstra}, \binits{R.}},
\bauthor{\bsnm{Rouder}, \binits{J.N.}},
\bauthor{\bsnm{Lee}, \binits{M.D.}},
\bauthor{\bsnm{Wagenmakers}, \binits{E.-J.}}:
\batitle{The fallacy of placing confidence in confidence intervals}.
\bjtitle{Psychonomic Bulletin \& Review}
\bvolume{23}(\bissue{1}),
\bfpage{103}--\blpage{123}
(\byear{2015})
\doiurl{10.3758/s13423-015-0947-8}
\end{barticle}
\endbibitem

\bibitem[\protect\citeauthoryear{Efron}{2012}]{efronBayesianInferenceParametric2012}
\begin{botherref}
\oauthor{\bsnm{Efron}, \binits{B.}}:
Bayesian inference and the parametric bootstrap.
The Annals of Applied Statistics
\textbf{6}(4)
(2012)
\doiurl{10.1214/12-AOAS571}
\end{botherref}
\endbibitem

\bibitem[\protect\citeauthoryear{Berger
  et~al.}{1994}]{bergerOverviewRobustBayesian1994}
\begin{barticle}
\bauthor{\bsnm{Berger}, \binits{J.O.}},
\bauthor{\bsnm{Moreno}, \binits{E.}},
\bauthor{\bsnm{Pericchi}, \binits{L.R.}},
\bauthor{\bsnm{Bayarri}, \binits{M.J.}},
\bauthor{\bsnm{Bernardo}, \binits{J.M.}},
\bauthor{\bsnm{Cano}, \binits{J.A.}},
\bauthor{\bsnm{{De la Horra}}, \binits{J.}},
\bauthor{\bsnm{Mart{\'i}n}, \binits{J.}},
\bauthor{\bsnm{{R{\'i}os-Ins{\'u}a}}, \binits{D.}},
\bauthor{\bsnm{Betr{\`o}}, \binits{B.}},
\bauthor{\bsnm{Dasgupta}, \binits{A.}},
\bauthor{\bsnm{Gustafson}, \binits{P.}},
\bauthor{\bsnm{Wasserman}, \binits{L.}},
\bauthor{\bsnm{Kadane}, \binits{J.B.}},
\bauthor{\bsnm{Srinivasan}, \binits{C.}},
\bauthor{\bsnm{Lavine}, \binits{M.}},
\bauthor{\bsnm{O'Hagan}, \binits{A.}},
\bauthor{\bsnm{Polasek}, \binits{W.}},
\bauthor{\bsnm{Robert}, \binits{C.P.}},
\bauthor{\bsnm{Goutis}, \binits{C.}},
\bauthor{\bsnm{Ruggeri}, \binits{F.}},
\bauthor{\bsnm{Salinetti}, \binits{G.}},
\bauthor{\bsnm{Sivaganesan}, \binits{S.}}:
\batitle{An overview of robust {{Bayesian}} analysis}.
\bjtitle{Test}
\bvolume{3}(\bissue{1}),
\bfpage{5}--\blpage{124}
(\byear{1994})
\doiurl{10.1007/BF02562676}
\end{barticle}
\endbibitem

\bibitem[\protect\citeauthoryear{Caesar
  et~al.}{2018}]{Caesar2018AMOCFingerprint}
\begin{barticle}
\bauthor{\bsnm{Caesar}, \binits{L.}},
\bauthor{\bsnm{Rahmstorf}, \binits{S.}},
\bauthor{\bsnm{Robinson}, \binits{A.}},
\bauthor{\bsnm{Feulner}, \binits{G.}},
\bauthor{\bsnm{Saba}, \binits{V.}}:
\batitle{Observed fingerprint of a weakening {Atlantic} {Ocean} overturning
  circulation}.
\bjtitle{Nature}
\bvolume{556}(\bissue{7700}),
\bfpage{191}--\blpage{196}
(\byear{2018})
\doiurl{10.1038/s41586-018-0006-5}
\end{barticle}
\endbibitem

\bibitem[\protect\citeauthoryear{Kanzow et~al.}{2010}]{Kanzow2010Seasonal}
\begin{barticle}
\bauthor{\bsnm{Kanzow}, \binits{T.}},
\bauthor{\bsnm{Cunningham}, \binits{S.A.}},
\bauthor{\bsnm{Johns}, \binits{W.E.}},
\bauthor{\bsnm{Hirschi}, \binits{J.J.-M.}},
\bauthor{\bsnm{Marotzke}, \binits{J.}},
\bauthor{\bsnm{Baringer}, \binits{M.O.}},
\bauthor{\bsnm{Meinen}, \binits{C.S.}},
\bauthor{\bsnm{Chidichimo}, \binits{M.P.}},
\bauthor{\bsnm{Atkinson}, \binits{C.}},
\bauthor{\bsnm{Beal}, \binits{L.M.}},
\bauthor{\bsnm{Bryden}, \binits{H.L.}},
\bauthor{\bsnm{Collins}, \binits{J.}}:
\batitle{Seasonal variability of the atlantic meridional overturning
  circulation at 26.5°n}.
\bjtitle{Journal of Climate}
\bvolume{23}(\bissue{21}),
\bfpage{5678}--\blpage{5698}
(\byear{2010})
\doiurl{10.1175/2010JCLI3389.1}
\end{barticle}
\endbibitem

\bibitem[\protect\citeauthoryear{Frajka-Williams
  et~al.}{2019}]{Frajka-Williams2019AtlanticVariability}
\begin{barticle}
\bauthor{\bsnm{Frajka-Williams}, \binits{E.}},
\bauthor{\bsnm{Ansorge}, \binits{I.J.}},
\bauthor{\bsnm{Baehr}, \binits{J.}},
\bauthor{\bsnm{Bryden}, \binits{H.L.}},
\bauthor{\bsnm{Chidichimo}, \binits{M.P.}},
\bauthor{\bsnm{Cunningham}, \binits{S.A.}},
\bauthor{\bsnm{Danabasoglu}, \binits{G.}},
\bauthor{\bsnm{Dong}, \binits{S.}},
\bauthor{\bsnm{Donohue}, \binits{K.A.}},
\bauthor{\bsnm{Elipot}, \binits{S.}},
\bauthor{\bsnm{Heimbach}, \binits{P.}},
\bauthor{\bsnm{Holliday}, \binits{N.P.}},
\bauthor{\bsnm{Hummels}, \binits{R.}},
\bauthor{\bsnm{Jackson}, \binits{L.C.}},
\bauthor{\bsnm{Karstensen}, \binits{J.}},
\bauthor{\bsnm{Lankhorst}, \binits{M.}},
\bauthor{\bsnm{Le~Bras}, \binits{I.A.}},
\bauthor{\bsnm{Susan~Lozier}, \binits{M.}},
\bauthor{\bsnm{McDonagh}, \binits{E.L.}},
\bauthor{\bsnm{Meinen}, \binits{C.S.}},
\bauthor{\bsnm{Mercier}, \binits{H.}},
\bauthor{\bsnm{Moat}, \binits{B.I.}},
\bauthor{\bsnm{Perez}, \binits{R.C.}},
\bauthor{\bsnm{Piecuch}, \binits{C.G.}},
\bauthor{\bsnm{Rhein}, \binits{M.}},
\bauthor{\bsnm{Srokosz}, \binits{M.A.}},
\bauthor{\bsnm{Trenberth}, \binits{K.E.}},
\bauthor{\bsnm{Bacon}, \binits{S.}},
\bauthor{\bsnm{Forget}, \binits{G.}},
\bauthor{\bsnm{Goni}, \binits{G.}},
\bauthor{\bsnm{Kieke}, \binits{D.}},
\bauthor{\bsnm{Koelling}, \binits{J.}},
\bauthor{\bsnm{Lamont}, \binits{T.}},
\bauthor{\bsnm{McCarthy}, \binits{G.D.}},
\bauthor{\bsnm{Mertens}, \binits{C.}},
\bauthor{\bsnm{Send}, \binits{U.}},
\bauthor{\bsnm{Smeed}, \binits{D.A.}},
\bauthor{\bsnm{Speich}, \binits{S.}},
\bauthor{\bsnm{Berg}, \binits{M.}},
\bauthor{\bsnm{Volkov}, \binits{D.}},
\bauthor{\bsnm{Wilson}, \binits{C.}}:
\batitle{{Atlantic meridional overturning circulation: Observed transport and
  variability}}.
\bjtitle{Frontiers in Marine Science}
\bvolume{6}(\bissue{JUN}),
\bfpage{1}--\blpage{18}
(\byear{2019})
\doiurl{10.3389/fmars.2019.00260}
\end{barticle}
\endbibitem

\bibitem[\protect\citeauthoryear{Frajka-Williams et~al.}{2021}]{RAPID}
\begin{botherref}
\oauthor{\bsnm{Frajka-Williams}, \binits{E.}},
\oauthor{\bsnm{Moat}, \binits{B.I.}},
\oauthor{\bsnm{Smeed}, \binits{D.A.}},
\oauthor{\bsnm{Rayner}, \binits{D.}},
\oauthor{\bsnm{Johns}, \binits{W.E.}},
\oauthor{\bsnm{Baringer}, \binits{M.O.}},
\oauthor{\bsnm{Volkov}, \binits{D.}},
\oauthor{\bsnm{Collins}, \binits{J.}}:
Atlantic meridional overturning circulation observed by the RAPID-MOCHA-WBTS
  (RAPID-Meridional Overturning Circulation and Heatflux Array-Western Boundary
  Time Series) array at 26N from 2004 to 2020 (v2020.1)
(2021).
\doiurl{10.5285/cc1e34b3-3385-662b-e053-6c86abc03444}
\end{botherref}
\endbibitem

\bibitem[\protect\citeauthoryear{Bryden et~al.}{2005}]{bryden2005Slowing}
\begin{barticle}
\bauthor{\bsnm{Bryden}, \binits{H.L.}},
\bauthor{\bsnm{Longworth}, \binits{H.R.}},
\bauthor{\bsnm{Cunningham}, \binits{S.A.}}:
\batitle{Slowing of the atlantic meridional overturning circulation at 25° n}.
\bjtitle{Nature}
\bvolume{438},
\bfpage{655}--\blpage{657}
(\byear{2005})
\doiurl{10.1038/nature04385}
\end{barticle}
\endbibitem

\bibitem[\protect\citeauthoryear{Rahmstorf
  et~al.}{2015}]{Rahmstorf2015ExceptionalCirculation}
\begin{barticle}
\bauthor{\bsnm{Rahmstorf}, \binits{S.}},
\bauthor{\bsnm{Box}, \binits{J.E.}},
\bauthor{\bsnm{Feulner}, \binits{G.}},
\bauthor{\bsnm{Mann}, \binits{M.E.}},
\bauthor{\bsnm{Robinson}, \binits{A.}},
\bauthor{\bsnm{Rutherford}, \binits{S.}},
\bauthor{\bsnm{Schaffernicht}, \binits{E.J.}}:
\batitle{{Exceptional twentieth-century slowdown in Atlantic Ocean overturning
  circulation}}.
\bjtitle{Nature Climate Change}
\bvolume{5}(\bissue{5}),
\bfpage{475}--\blpage{480}
(\byear{2015})
\doiurl{10.1038/nclimate2554}
\end{barticle}
\endbibitem

\bibitem[\protect\citeauthoryear{Caesar
  et~al.}{2018}]{Caesar2018ObservedCirculation}
\begin{barticle}
\bauthor{\bsnm{Caesar}, \binits{L.}},
\bauthor{\bsnm{Rahmstorf}, \binits{S.}},
\bauthor{\bsnm{Robinson}, \binits{A.}},
\bauthor{\bsnm{Feulner}, \binits{G.}},
\bauthor{\bsnm{Saba}, \binits{V.}}:
\batitle{{Observed fingerprint of a weakening Atlantic Ocean overturning
  circulation}}.
\bjtitle{Nature}
\bvolume{556}(\bissue{7700}),
\bfpage{191}--\blpage{196}
(\byear{2018})
\doiurl{10.1038/s41586-018-0006-5}
\end{barticle}
\endbibitem

\bibitem[\protect\citeauthoryear{Drijfhout
  et~al.}{2012}]{Drijfhout2012IsPatterns}
\begin{barticle}
\bauthor{\bsnm{Drijfhout}, \binits{S.}},
\bauthor{\bsnm{Oldenborgh}, \binits{G.J.}},
\bauthor{\bsnm{Cimatoribus}, \binits{A.}}:
\batitle{{Is a decline of AMOC causing the warming hole above the North
  Atlantic in observed and modeled warming patterns?}}
\bjtitle{Journal of Climate}
\bvolume{25}(\bissue{24}),
\bfpage{8373}--\blpage{8379}
(\byear{2012})
\doiurl{10.1175/JCLI-D-12-00490.1}
\end{barticle}
\endbibitem

\bibitem[\protect\citeauthoryear{Menary and Wood}{2018}]{Menary2018AnModels}
\begin{barticle}
\bauthor{\bsnm{Menary}, \binits{M.B.}},
\bauthor{\bsnm{Wood}, \binits{R.A.}}:
\batitle{{An anatomy of the projected North Atlantic warming hole in CMIP5
  models}}.
\bjtitle{Climate Dynamics}
\bvolume{50}(\bissue{7-8}),
\bfpage{3063}--\blpage{3080}
(\byear{2018})
\doiurl{10.1007/s00382-017-3793-8}
\end{barticle}
\endbibitem

\bibitem[\protect\citeauthoryear{Liu et~al.}{2020}]{Liu2020ClimateClimate}
\begin{barticle}
\bauthor{\bsnm{Liu}, \binits{W.}},
\bauthor{\bsnm{Fedorov}, \binits{A.V.}},
\bauthor{\bsnm{Xie}, \binits{S.P.}},
\bauthor{\bsnm{Hu}, \binits{S.}}:
\batitle{{Climate impacts of a weakened Atlantic meridional overturning
  circulation in a warming climate}}.
\bjtitle{Science Advances}
\bvolume{6}(\bissue{26}),
\bfpage{2}--\blpage{10}
(\byear{2020})
\doiurl{10.1126/sciadv.aaz4876}
\end{barticle}
\endbibitem

\bibitem[\protect\citeauthoryear{Little et~al.}{2020}]{Little2020DoStrength}
\begin{barticle}
\bauthor{\bsnm{Little}, \binits{C.M.}},
\bauthor{\bsnm{Zhao}, \binits{M.}},
\bauthor{\bsnm{Buckley}, \binits{M.W.}}:
\batitle{{Do Surface Temperature Indices Reflect Centennial-Timescale Trends in
  Atlantic Meridional Overturning Circulation Strength?}}
\bjtitle{Geophysical Research Letters}
\bvolume{47}(\bissue{22}),
\bfpage{1}--\blpage{10}
(\byear{2020})
\doiurl{10.1029/2020GL090888}
\end{barticle}
\endbibitem

\bibitem[\protect\citeauthoryear{Jackson and
  Wood}{2018}]{Jackson2018HysteresisGCM}
\begin{barticle}
\bauthor{\bsnm{Jackson}, \binits{L.C.}},
\bauthor{\bsnm{Wood}, \binits{R.A.}}:
\batitle{{Hysteresis and Resilience of the AMOC in an Eddy-Permitting GCM}}.
\bjtitle{Geophysical Research Letters}
\bvolume{45}(\bissue{16}),
\bfpage{8547}--\blpage{8556}
(\byear{2018})
\doiurl{10.1029/2018GL078104}
\end{barticle}
\endbibitem

\bibitem[\protect\citeauthoryear{Li
  et~al.}{2022}]{Li2022Century-longExplanation}
\begin{barticle}
\bauthor{\bsnm{Li}, \binits{L.}},
\bauthor{\bsnm{Lozier}, \binits{M.S.}},
\bauthor{\bsnm{Li}, \binits{F.}}:
\batitle{{Century-long cooling trend in subpolar North Atlantic forced by
  atmosphere: an alternative explanation}}.
\bjtitle{Climate Dynamics}
\bvolume{58}(\bissue{9-10}),
\bfpage{2249}--\blpage{2267}
(\byear{2022})
\doiurl{10.1007/s00382-021-06003-4}
\end{barticle}
\endbibitem

\bibitem[\protect\citeauthoryear{He et~al.}{2022}]{He2022ACirculation}
\begin{botherref}
\oauthor{\bsnm{He}, \binits{C.}},
\oauthor{\bsnm{Clement}, \binits{A.C.}},
\oauthor{\bsnm{Cane}, \binits{M.A.}},
\oauthor{\bsnm{Murphy}, \binits{L.N.}},
\oauthor{\bsnm{Klavans}, \binits{J.M.}},
\oauthor{\bsnm{Fenske}, \binits{T.M.}}:
{A North Atlantic Warming Hole Without Ocean Circulation}.
Geophysical Research Letters
\textbf{49}(19)
(2022)
\doiurl{10.1029/2022GL100420}
\end{botherref}
\endbibitem

\bibitem[\protect\citeauthoryear{Ferster
  et~al.}{2022}]{Ferster2022SlowdownDecline}
\begin{botherref}
\oauthor{\bsnm{Ferster}, \binits{B.S.}},
\oauthor{\bsnm{Simon}, \binits{A.}},
\oauthor{\bsnm{Fedorov}, \binits{A.}},
\oauthor{\bsnm{Mignot}, \binits{J.}},
\oauthor{\bsnm{Guilyardi}, \binits{E.}}:
{Slowdown and Recovery of the Atlantic Meridional Overturning Circulation and a
  Persistent North Atlantic Warming Hole Induced by Arctic Sea Ice Decline}.
Geophysical Research Letters
\textbf{49}(16)
(2022)
\doiurl{10.1029/2022GL097967}
\end{botherref}
\endbibitem

\bibitem[\protect\citeauthoryear{Ghosh et~al.}{2022}]{Ghosh2022TwoWarming}
\begin{barticle}
\bauthor{\bsnm{Ghosh}, \binits{R.}},
\bauthor{\bsnm{Putrasahan}, \binits{D.}},
\bauthor{\bsnm{Manzini}, \binits{E.}},
\bauthor{\bsnm{Lohmann}, \binits{K.}},
\bauthor{\bsnm{Keil}, \binits{P.}},
\bauthor{\bsnm{Hand}, \binits{R.}},
\bauthor{\bsnm{Bader}, \binits{J.}},
\bauthor{\bsnm{Matei}, \binits{D.}},
\bauthor{\bsnm{Jungclaus}, \binits{J.H.}}:
\batitle{{Two distinct phases of North Atlantic Eastern Subpolar Gyre and
  Warming Hole evolution under Global Warming}}.
\bjtitle{Journal of Climate}
\bvolume{36}(\bissue{6}),
\bfpage{1}--\blpage{31}
(\byear{2022})
\doiurl{10.1175/JCLI-D-22-0222.1.}
\end{barticle}
\endbibitem

\bibitem[\protect\citeauthoryear{Keil et~al.}{2020}]{Keil2020MultipleHole}
\begin{barticle}
\bauthor{\bsnm{Keil}, \binits{P.}},
\bauthor{\bsnm{Mauritsen}, \binits{T.}},
\bauthor{\bsnm{Jungclaus}, \binits{J.}},
\bauthor{\bsnm{Hedemann}, \binits{C.}},
\bauthor{\bsnm{Olonscheck}, \binits{D.}},
\bauthor{\bsnm{Ghosh}, \binits{R.}}:
\batitle{{Multiple drivers of the North Atlantic warming hole}}.
\bjtitle{Nature Climate Change}
\bvolume{10}(\bissue{7}),
\bfpage{667}--\blpage{671}
(\byear{2020})
\doiurl{10.1038/s41558-020-0819-8}
\end{barticle}
\endbibitem

\bibitem[\protect\citeauthoryear{Lozier et~al.}{2019}]{Lozier2019AAtlantic}
\begin{barticle}
\bauthor{\bsnm{Lozier}, \binits{M.S.}},
\bauthor{\bsnm{Li}, \binits{F.}},
\bauthor{\bsnm{Bacon}, \binits{S.}},
\bauthor{\bsnm{Bahr}, \binits{F.}},
\bauthor{\bsnm{Bower}, \binits{A.S.}},
\bauthor{\bsnm{Cunningham}, \binits{S.A.}},
\bauthor{\bsnm{De~Jong}, \binits{M.F.}},
\bauthor{\bsnm{De~Steur}, \binits{L.}},
\bauthor{\bsnm{DeYoung}, \binits{B.}},
\bauthor{\bsnm{Fischer}, \binits{J.}},
\bauthor{\bsnm{Gary}, \binits{S.F.}},
\bauthor{\bsnm{Greenan}, \binits{B.J.W.}},
\bauthor{\bsnm{Holliday}, \binits{N.P.}},
\bauthor{\bsnm{Houk}, \binits{A.}},
\bauthor{\bsnm{Houpert}, \binits{L.}},
\bauthor{\bsnm{Inall}, \binits{M.E.}},
\bauthor{\bsnm{Johns}, \binits{W.E.}},
\bauthor{\bsnm{Johnson}, \binits{H.L.}},
\bauthor{\bsnm{Johnson}, \binits{C.}},
\bauthor{\bsnm{Karstensen}, \binits{J.}},
\bauthor{\bsnm{Koman}, \binits{G.}},
\bauthor{\bsnm{Le~Bras}, \binits{I.A.}},
\bauthor{\bsnm{Lin}, \binits{X.}},
\bauthor{\bsnm{Mackay}, \binits{N.}},
\bauthor{\bsnm{Marshall}, \binits{D.P.}},
\bauthor{\bsnm{Mercier}, \binits{H.}},
\bauthor{\bsnm{Oltmanns}, \binits{M.}},
\bauthor{\bsnm{Pickart}, \binits{R.S.}},
\bauthor{\bsnm{Ramsey}, \binits{A.L.}},
\bauthor{\bsnm{Rayner}, \binits{D.}},
\bauthor{\bsnm{Straneo}, \binits{F.}},
\bauthor{\bsnm{Thierry}, \binits{V.}},
\bauthor{\bsnm{Torres}, \binits{D.J.}},
\bauthor{\bsnm{Williams}, \binits{R.G.}},
\bauthor{\bsnm{Wilson}, \binits{C.}},
\bauthor{\bsnm{Yang}, \binits{J.}},
\bauthor{\bsnm{Yashayaev}, \binits{I.}},
\bauthor{\bsnm{Zhao}, \binits{J.}}:
\batitle{{A sea change in our view of overturning in the subpolar North
  Atlantic}}.
\bjtitle{Science}
\bvolume{363}(\bissue{6426}),
\bfpage{516}--\blpage{521}
(\byear{2019})
\doiurl{10.1126/science.aau6592}
\end{barticle}
\endbibitem

\bibitem[\protect\citeauthoryear{Chafik
  et~al.}{2022}]{Chafik2022IrmingerVariability}
\begin{barticle}
\bauthor{\bsnm{Chafik}, \binits{L.}},
\bauthor{\bsnm{Holliday}, \binits{N.P.}},
\bauthor{\bsnm{Bacon}, \binits{S.}},
\bauthor{\bsnm{Rossby}, \binits{T.}}:
\batitle{{Irminger Sea Is the Center of Action for Subpolar AMOC Variability}}.
\bjtitle{Geophysical Research Letters}
\bvolume{49}(\bissue{17}),
\bfpage{1}--\blpage{11}
(\byear{2022})
\doiurl{10.1029/2022GL099133}
\end{barticle}
\endbibitem

\bibitem[\protect\citeauthoryear{Sgubin et~al.}{2017}]{Sgubin2017AbruptModels}
\begin{botherref}
\oauthor{\bsnm{Sgubin}, \binits{G.}},
\oauthor{\bsnm{Swingedouw}, \binits{D.}},
\oauthor{\bsnm{Drijfhout}, \binits{S.}},
\oauthor{\bsnm{Mary}, \binits{Y.}},
\oauthor{\bsnm{Bennabi}, \binits{A.}}:
{Abrupt cooling over the North Atlantic in modern climate models}.
Nature Communications
\textbf{8}
(2017)
\doiurl{10.1038/ncomms14375}
\end{botherref}
\endbibitem

\bibitem[\protect\citeauthoryear{Swingedouw
  et~al.}{2021}]{Swingedouw2021OnModels}
\begin{barticle}
\bauthor{\bsnm{Swingedouw}, \binits{D.}},
\bauthor{\bsnm{Bily}, \binits{A.}},
\bauthor{\bsnm{Esquerdo}, \binits{C.}},
\bauthor{\bsnm{Borchert}, \binits{L.F.}},
\bauthor{\bsnm{Sgubin}, \binits{G.}},
\bauthor{\bsnm{Mignot}, \binits{J.}},
\bauthor{\bsnm{Menary}, \binits{M.}}:
\batitle{{On the risk of abrupt changes in the North Atlantic subpolar gyre in
  CMIP6 models}}.
\bjtitle{Annals of the New York Academy of Sciences}
\bvolume{1504}(\bissue{1}),
\bfpage{187}--\blpage{201}
(\byear{2021})
\doiurl{10.1111/nyas.14659}
\end{barticle}
\endbibitem

\bibitem[\protect\citeauthoryear{McKay et~al.}{2022}]{McKay2022ExceedingPoints}
\begin{botherref}
\oauthor{\bsnm{McKay}, \binits{D.I.A.}},
\oauthor{\bsnm{Staal}, \binits{A.}},
\oauthor{\bsnm{Abrams}, \binits{J.F.}},
\oauthor{\bsnm{Winkelmann}, \binits{R.}},
\oauthor{\bsnm{Sakschewski}, \binits{B.}},
\oauthor{\bsnm{Loriani}, \binits{S.}},
\oauthor{\bsnm{Fetzer}, \binits{I.}},
\oauthor{\bsnm{Cornell}, \binits{S.E.}},
\oauthor{\bsnm{Rockstr{\"{o}}m}, \binits{J.}},
\oauthor{\bsnm{Lenton}, \binits{T.M.}}:
{Exceeding 1.5{${}^\circ$}C global warming could trigger multiple climate
  tipping points}.
Science
\textbf{377}(6611)
(2022)
\doiurl{10.1126/science.abn7950}
\end{botherref}
\endbibitem

\bibitem[\protect\citeauthoryear{Ben-Yami
  et~al.}{2023}]{BenYami2023AMOCDataCSD}
\begin{barticle}
\bauthor{\bsnm{Ben-Yami}, \binits{M.}},
\bauthor{\bsnm{Skiba}, \binits{V.}},
\bauthor{\bsnm{Bathiany}, \binits{S.}},
\bauthor{\bsnm{Boers}, \binits{N.}}:
\batitle{Uncertainties in critical slowing down indicators of observation-based
  fingerprints of the {Atlantic} {Overturning} {Circulation}}.
\bjtitle{Nature Communications}
\bvolume{14}(\bissue{1}),
\bfpage{8344}
(\byear{2023})
\doiurl{10.1038/s41467-023-44046-9}
\end{barticle}
\endbibitem

\bibitem[\protect\citeauthoryear{Roberts
  et~al.}{2013}]{Roberts2013ACirculation}
\begin{barticle}
\bauthor{\bsnm{Roberts}, \binits{C.D.}},
\bauthor{\bsnm{Garry}, \binits{F.K.}},
\bauthor{\bsnm{Jackson}, \binits{L.C.}}:
\batitle{{A multimodel study of sea surface temperature and subsurface density
  fingerprints of the Atlantic meridional overturning circulation}}.
\bjtitle{Journal of Climate}
\bvolume{26}(\bissue{22}),
\bfpage{9155}--\blpage{9174}
(\byear{2013})
\doiurl{10.1175/JCLI-D-12-00762.1}
\end{barticle}
\endbibitem

\bibitem[\protect\citeauthoryear{Jackson and
  Wood}{2020}]{Jackson2020FingerprintsAMOC}
\begin{barticle}
\bauthor{\bsnm{Jackson}, \binits{L.C.}},
\bauthor{\bsnm{Wood}, \binits{R.A.}}:
\batitle{Fingerprints for {Early} {Detection} of {Changes} in the {AMOC}}.
\bjtitle{Journal of Climate}
\bvolume{33}(\bissue{16}),
\bfpage{7027}--\blpage{7044}
(\byear{2020})
\doiurl{10.1175/JCLI-D-20-0034.1}
\end{barticle}
\endbibitem

\bibitem[\protect\citeauthoryear{Smith
  et~al.}{2023}]{Smith2023ReliabilitySeries}
\begin{barticle}
\bauthor{\bsnm{Smith}, \binits{T.}},
\bauthor{\bsnm{Zotta}, \binits{R.M.}},
\bauthor{\bsnm{Boulton}, \binits{C.A.}},
\bauthor{\bsnm{Lenton}, \binits{T.M.}},
\bauthor{\bsnm{Dorigo}, \binits{W.}},
\bauthor{\bsnm{Boers}, \binits{N.}}:
\batitle{{Reliability of resilience estimation based on multi-instrument time
  series}}.
\bjtitle{Earth System Dynamics}
\bvolume{14}(\bissue{1}),
\bfpage{173}--\blpage{183}
(\byear{2023})
\doiurl{10.5194/esd-14-173-2023}
\end{barticle}
\endbibitem

\bibitem[\protect\citeauthoryear{Boers}{2023}]{Boers2023ReplyCirculation}
\begin{barticle}
\bauthor{\bsnm{Boers}, \binits{N.}}:
\batitle{{Reply to: Evidence lacking for a pending collapse of the Atlantic
  Meridional Overturning Circulation}}.
\bjtitle{Nature Climate Change}
\bvolume{14}(\bissue{January}),
\bfpage{43}--\blpage{48}
(\byear{2023})
\doiurl{10.1038/s41558-023-01878-z}
\end{barticle}
\endbibitem

\bibitem[\protect\citeauthoryear{Lundstad et~al.}{2023}]{Lundstad2023TheHCLIM}
\begin{barticle}
\bauthor{\bsnm{Lundstad}, \binits{E.}},
\bauthor{\bsnm{Brugnara}, \binits{Y.}},
\bauthor{\bsnm{Pappert}, \binits{D.}},
\bauthor{\bsnm{Kopp}, \binits{J.}},
\bauthor{\bsnm{Samakinwa}, \binits{E.}},
\bauthor{\bsnm{H{\"{u}}rzeler}, \binits{A.}},
\bauthor{\bsnm{Andersson}, \binits{A.}},
\bauthor{\bsnm{Chimani}, \binits{B.}},
\bauthor{\bsnm{Cornes}, \binits{R.}},
\bauthor{\bsnm{Demar{\'{e}}e}, \binits{G.}},
\bauthor{\bsnm{Filipiak}, \binits{J.}},
\bauthor{\bsnm{Gates}, \binits{L.}},
\bauthor{\bsnm{Ives}, \binits{G.L.}},
\bauthor{\bsnm{Jones}, \binits{J.M.}},
\bauthor{\bsnm{Jourdain}, \binits{S.}},
\bauthor{\bsnm{Kiss}, \binits{A.}},
\bauthor{\bsnm{Nicholson}, \binits{S.E.}},
\bauthor{\bsnm{Przybylak}, \binits{R.}},
\bauthor{\bsnm{Jones}, \binits{P.}},
\bauthor{\bsnm{Rousseau}, \binits{D.}},
\bauthor{\bsnm{Tinz}, \binits{B.}},
\bauthor{\bsnm{Rodrigo}, \binits{F.S.}},
\bauthor{\bsnm{Grab}, \binits{S.}},
\bauthor{\bsnm{Dom{\'{i}}nguez-Castro}, \binits{F.}},
\bauthor{\bsnm{Slonosky}, \binits{V.}},
\bauthor{\bsnm{Cooper}, \binits{J.}},
\bauthor{\bsnm{Brunet}, \binits{M.}},
\bauthor{\bsnm{Br{\"{o}}nnimann}, \binits{S.}}:
\batitle{{The global historical climate database HCLIM}}.
\bjtitle{Scientific Data}
\bvolume{10}(\bissue{1}),
\bfpage{1}--\blpage{16}
(\byear{2023})
\doiurl{10.1038/s41597-022-01919-w}
\end{barticle}
\endbibitem

\bibitem[\protect\citeauthoryear{Kennedy et~al.}{2019}]{Kennedy2019AnSet}
\begin{barticle}
\bauthor{\bsnm{Kennedy}, \binits{J.J.}},
\bauthor{\bsnm{Rayner}, \binits{N.A.}},
\bauthor{\bsnm{Atkinson}, \binits{C.P.}},
\bauthor{\bsnm{Killick}, \binits{R.E.}}:
\batitle{{An Ensemble Data Set of Sea Surface Temperature Change From 1850: The
  Met Office Hadley Centre HadSST.4.0.0.0 Data Set}}.
\bjtitle{Journal of Geophysical Research: Atmospheres}
\bvolume{124}(\bissue{14}),
\bfpage{7719}--\blpage{7763}
(\byear{2019})
\doiurl{10.1029/2018JD029867}
\end{barticle}
\endbibitem

\bibitem[\protect\citeauthoryear{Rayner et~al.}{2003}]{Rayner2003GlobalCentury}
\begin{botherref}
\oauthor{\bsnm{Rayner}, \binits{N.A.}},
\oauthor{\bsnm{Parker}, \binits{D.E.}},
\oauthor{\bsnm{Horton}, \binits{E.B.}},
\oauthor{\bsnm{Folland}, \binits{C.K.}},
\oauthor{\bsnm{Alexander}, \binits{L.V.}},
\oauthor{\bsnm{Rowell}, \binits{D.P.}},
\oauthor{\bsnm{Kent}, \binits{E.C.}},
\oauthor{\bsnm{Kaplan}, \binits{A.}}:
{Global analyses of sea surface temperature, sea ice, and night marine air
  temperature since the late nineteenth century}.
Journal of Geophysical Research: Atmospheres
\textbf{108}(14)
(2003)
\doiurl{10.1029/2002jd002670}
\end{botherref}
\endbibitem

\bibitem[\protect\citeauthoryear{Smith
  et~al.}{2023}]{Smith2023ReliabilityResilienceEstimation}
\begin{barticle}
\bauthor{\bsnm{Smith}, \binits{T.}},
\bauthor{\bsnm{Zotta}, \binits{R.-M.}},
\bauthor{\bsnm{Boulton}, \binits{C.A.}},
\bauthor{\bsnm{Lenton}, \binits{T.M.}},
\bauthor{\bsnm{Dorigo}, \binits{W.}},
\bauthor{\bsnm{Boers}, \binits{N.}}:
\batitle{Reliability of resilience estimation based on multi-instrument time
  series}.
\bjtitle{Earth System Dynamics}
\bvolume{14}(\bissue{1}),
\bfpage{173}--\blpage{183}
(\byear{2023})
\doiurl{10.5194/esd-14-173-2023}
\end{barticle}
\endbibitem

\bibitem[\protect\citeauthoryear{Huang
  et~al.}{2017}]{Huang2017ExtendedIntercomparisons}
\begin{barticle}
\bauthor{\bsnm{Huang}, \binits{B.}},
\bauthor{\bsnm{Thorne}, \binits{P.W.}},
\bauthor{\bsnm{Banzon}, \binits{V.F.}},
\bauthor{\bsnm{Boyer}, \binits{T.}},
\bauthor{\bsnm{Chepurin}, \binits{G.}},
\bauthor{\bsnm{Lawrimore}, \binits{J.H.}},
\bauthor{\bsnm{Menne}, \binits{M.J.}},
\bauthor{\bsnm{Smith}, \binits{T.M.}},
\bauthor{\bsnm{Vose}, \binits{R.S.}},
\bauthor{\bsnm{Zhang}, \binits{H.M.}}:
\batitle{{Extended reconstructed Sea surface temperature, Version 5 (ERSSTv5):
  Upgrades, validations, and intercomparisons}}.
\bjtitle{Journal of Climate}
\bvolume{30}(\bissue{20}),
\bfpage{8179}--\blpage{8205}
(\byear{2017})
\doiurl{10.1175/JCLI-D-16-0836.1}
\end{barticle}
\endbibitem

\bibitem[\protect\citeauthoryear{Morice et~al.}{2021}]{Morice2021AnSet}
\begin{barticle}
\bauthor{\bsnm{Morice}, \binits{C.P.}},
\bauthor{\bsnm{Kennedy}, \binits{J.J.}},
\bauthor{\bsnm{Rayner}, \binits{N.A.}},
\bauthor{\bsnm{Winn}, \binits{J.P.}},
\bauthor{\bsnm{Hogan}, \binits{E.}},
\bauthor{\bsnm{Killick}, \binits{R.E.}},
\bauthor{\bsnm{Dunn}, \binits{R.J.H.}},
\bauthor{\bsnm{Osborn}, \binits{T.J.}},
\bauthor{\bsnm{Jones}, \binits{P.D.}},
\bauthor{\bsnm{Simpson}, \binits{I.R.}}:
\batitle{{An Updated Assessment of Near-Surface Temperature Change From 1850:
  The HadCRUT5 Data Set}}.
\bjtitle{Journal of Geophysical Research: Atmospheres}
\bvolume{126}(\bissue{3}),
\bfpage{1}--\blpage{28}
(\byear{2021})
\doiurl{10.1029/2019JD032361}
\end{barticle}
\endbibitem

\bibitem[\protect\citeauthoryear{Pilipovic
  et~al.}{2024}]{Pilipovic2024SDEParameterEstimation}
\begin{botherref}
\oauthor{\bsnm{Pilipovic}, \binits{P.}},
\oauthor{\bsnm{Samson}, \binits{A.}},
\oauthor{\bsnm{Ditlevsen}, \binits{S.}}:
Parameter estimation in nonlinear multivariate stochastic differential
  equations based on splitting schemes.
The Annals of Statistics
\textbf{52}(2)
(2024)
\doiurl{10.1214/24-AOS2371}
\end{botherref}
\endbibitem

\bibitem[\protect\citeauthoryear{{ter Braak} and
  Vrugt}{2008}]{terbraakDifferentialEvolutionMarkov2008}
\begin{barticle}
\bauthor{\bsnm{{ter Braak}}, \binits{C.J.F.}},
\bauthor{\bsnm{Vrugt}, \binits{J.A.}}:
\batitle{Differential {{Evolution Markov Chain}} with snooker updater and~fewer
  chains}.
\bjtitle{Statistics and Computing}
\bvolume{18}(\bissue{4}),
\bfpage{435}--\blpage{446}
(\byear{2008})
\doiurl{10.1007/s11222-008-9104-9}
\end{barticle}
\endbibitem

\bibitem[\protect\citeauthoryear{Hartig
  et~al.}{2017}]{hartigBayesianToolsGeneralPurposeMCMC2017}
\begin{botherref}
\oauthor{\bsnm{Hartig}, \binits{F.}},
\oauthor{\bsnm{Minunno}, \binits{F.}},
\oauthor{\bsnm{Paul}, \binits{S.}}:
{{BayesianTools}}: {{General-Purpose MCMC}} and {{SMC Samplers}} and {{Tools}}
  for {{Bayesian Statistics}}.
Comprehensive R Archive Network
(2017).
\doiurl{10.32614/CRAN.package.BayesianTools}
\end{botherref}
\endbibitem

\bibitem[\protect\citeauthoryear{Pérez-Hernández
  et~al.}{2023}]{Perez-Hernandez2023TheGyre}
\begin{barticle}
\bauthor{\bsnm{Pérez-Hernández}, \binits{M.D.}},
\bauthor{\bsnm{Hernández-Guerra}, \binits{A.}},
\bauthor{\bsnm{Cana-Cascallar}, \binits{L.}},
\bauthor{\bsnm{Arumí-Planas}, \binits{C.}},
\bauthor{\bsnm{Caínzos}, \binits{V.}},
\bauthor{\bsnm{González-Santana}, \binits{A.J.}},
\bauthor{\bsnm{Gutiérrez-Guerra}, \binits{M.A.}},
\bauthor{\bsnm{Martínez-Marrero}, \binits{A.}},
\bauthor{\bsnm{Mosquera~Giménez}, \binits{A.}},
\bauthor{\bsnm{Presas~Navarro}, \binits{C.}},
\bauthor{\bsnm{Santana-Toscano}, \binits{D.}},
\bauthor{\bsnm{Vélez-Belchí}, \binits{P.}}:
\batitle{The seasonal cycle of the eastern boundary currents of the north
  atlantic subtropical gyre}.
\bjtitle{Journal of Geophysical Research: Oceans}
\bvolume{128}(\bissue{4}),
\bfpage{2022}--\blpage{019487}
(\byear{2023})
\doiurl{10.1029/2022JC019487}
{\href{https://arxiv.org/abs/https://agupubs.onlinelibrary.wiley.com/doi/pdf/10.1029/2022JC019487}{{https://agupubs.onlinelibrary.wiley.com/doi/pdf/10.1029/2022JC019487}}}.
\bcomment{e2022JC019487 2022JC019487}
\end{barticle}
\endbibitem

\end{thebibliography}

\section*{Methods}

To assess the robustness of the tipping time predictions presented in DD23, we applied their Maximum Likelihood Estimation (MLE) method to a suite of synthetically generated time series, and combinations of alternative SST-based AMOC fingerprints and observational datasets. We additionally extend the likelihood formulation by replacing the quadratic drift term in Eq.~\eqref{eq: quad} with a more general third-order polynomial representation of the deterministic dynamics, introduced in Eq.~\eqref{eq: cubic} below.

\subsection*{Third-order drift extension of the maximum likelihood method}

We implemented an alternative statistical model within the DD23 maximum-likelihood framework by replacing the quadratic fold normal form drift with a cubic polynomial, following the formulation introduced in Cotronei et al.~\cite{Cotronei2026AMOCTippingUncertainty}. 
Instead of the DD23 stochastic differential equation (SDE)
\begin{equation}\label{eq: quad}
    \dint X_t = -\big(A (X_t - m)^2 + \lambda(t) \big)\, \dint t + \sigma\,\dint B_t,
\end{equation}
we used a model with a cubic deterministic drift
\begin{equation}\label{eq: cubic}
    \dint X_t = -\Big( A (X_t - m)^3 + C (X_t - m) + \lambda(t) \Big)\, \dint t + \sigma\,\dint B_t,
\end{equation}
which corresponds to Eq.~(12) in \cite{Cotronei2026AMOCTippingUncertainty}. 
Here $X_t$ denotes the AMOC fingerprint, $m$ a constant reference state, $A$ and $C$ shape the deterministic equilibrium branch, $\lambda_t = \lambda_0 (t - t_0)$ is the externally forced control parameter, and $\sigma$ is the noise amplitude. 
In contrast to the original drift used by DD23, this formulation does not strictly impose the existence of a fold bifurcation: depending on the fitted parameters, the equilibrium branch may exhibit either a fold or a smooth inflection.

Parameter estimation was performed by maximum likelihood using analogous discretized transition densities and the same numerical optimization procedure as in DD23. 
The likelihood was computed analogously to the ansatz of DD23 using Strang splitting of the SDE \cite{Pilipovic2024SDEParameterEstimation}, and the free parameters $(A=0.29, C=1.68, \tau_r=137.49)$ were estimated by minimizing the negative log-likelihood over the observed SPG-SST-based AMOC time series. For more information on the exact routine, see the Supplementary Material of Cotronei et al.~\cite{Cotronei2026AMOCTippingUncertainty}.
The time of onset of the external forcing, $t_0=1924$, was kept identical to DD23.

\subsection*{Synthetic time series from the model with third-order drift}

To test the sensitivity of the DD23 maximum-likelihood framework to the model choice, we generated synthetic time series from the stochastic model with cubic drift defined in Eq.~\eqref{eq: cubic}, using the parameter values inferred from the application of the MLE method with cubic drift to the observed AMOC fingerprint. 
This setup ensures that the synthetic data are statistically consistent with the observations, while originating from dynamics that do not necessarily contain a tipping point.

We generated 1000 independent realizations using an Euler-Maruyama discretization with twenty integration time steps per month.
Each realization was subsequently coarse-grained to observational resolution and analyzed using the original DD23 MLE procedure, without modification. 
This constitutes a controlled experiment in which the data-generating process is known, but the inference model is structurally misspecified. Following DD23's approach, we also tested whether the synthetic time series showed statistically significant differences in variance and lag-1 autocorrelation from their baseline values at the $5\%$ level. This ensures that the time series would not be rejected based on non-significant critical slowing down indicators (for more details, see the middle of page 4 in DD23). Approximately $75\%$ of model-generated realizations passed this criterion, identical to the rate seen for the model proposed by DD23.

\subsection*{Analysis of the bootstrap confidence interval procedure}
\begin{figure}
    \centering
    \includegraphics[width=\linewidth]{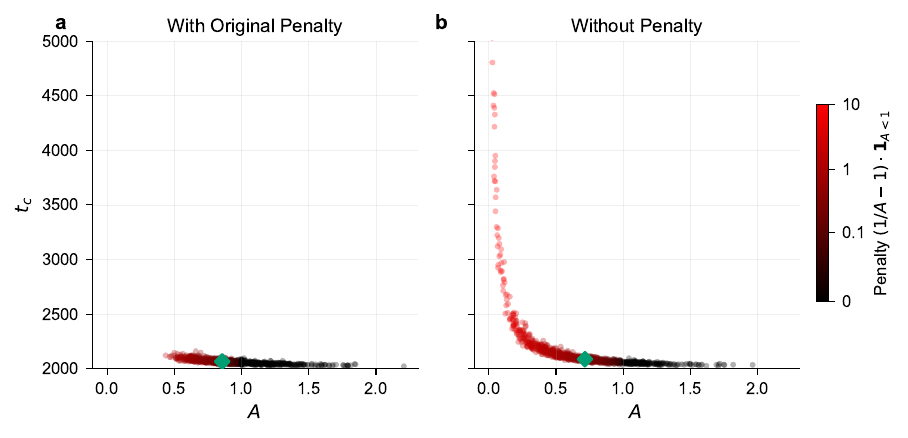}
    \caption{Maximum likelihood estimates of the parameters $A$ and $t_c$ obtained during the parametric bootstrap confidence interval procedure of DD23, applied to their AMOC fingerprint. Each dot corresponds to one of the 1000 synthetic bootstrap samples; the green diamond marks the MLE on the original observations. Panel (a) shows results with the original penalty on $A$, and panel (b) without it. Dot colour indicates the penalty value $(1/A - 1)\cdot\mathbf{1}_{A<1}$ at the estimated $A$. The penalty effectively suppresses large $t_c$ estimates by discouraging values of $A$ far below unity. High estimates with $t_c > 5000$ (reaching up to 30,000) are not depicted here.}
    \label{fig:AvsTc}
\end{figure}

A plausible explanation of the failure of the CI procedure used in DD23 with respect to the claimed 95\% coverage probability is the use of an additive penalty $p\cdot (\frac1A-1)\cdot\mathbf{1}_{A<1}$ on the approximate likelihood used in the MLE. The penalty is only mentioned in the supplementary information, where the authors justify its use by stating: \textit{Since division by $A$ enters the calculations of $\lambda_0$ and $m$ and thus the pseudo-likelihood, estimates are sensitive to small values of $A$. We therefore regularize the optimization problem by adding a penalization term on small values of $A$.} We do not see why this would justify the use of a penalty as the MLE is invariant under transformations such as $B=\frac1A$ or $B = \log A$ and the MLE based on $B$ does not seem to justify the equivalent penalty $p\cdot (B-1)\cdot\mathbf{1}_{B>1}$. As there is empirically a negative dependence between $A$ and $t_c$ (Figure \ref{fig:AvsTc}), the penalty introduces a bias towards smaller values of $t_c$ in the MLE. The reported CIs are thus doubly biased by the penalty: the biased MLE estimates are used to specify the data-generating process from which bootstrap samples are drawn, and the penalized MLE is then applied to each bootstrap sample to obtain the bootstrap distribution of $t_c$, thereby amplifying the bias.

Moreover, the data-driven choices of $t_0$ and $p$ are not accounted for in the bootstrap procedure that produces the CI. The penalty weight $p$ in particular is highly influential on the results, as shown in Figure \ref{fig:coverageSimulation}g. Specifically, $p$ is chosen as follows: first, apply the unpenalized MLE; then simulate many (we use 1000) trajectories from the SDE Equation \eqref{eq: DD23} using the resulting parameter estimates; finally, optimize $p$ over a grid of values to minimize the mean squared error between the true ramping time $\tau_r$ and its penalized-MLE estimate. This selection procedure is not documented in DD23 or in the published code, where $p$ is hardcoded to $0.004$. The supplementary information only states that $p$ was \textit{determined by cross-validation on simulated data sets by minimizing the mean squared distance between the estimated ramping time on each data set and the value of the ramping time used in the simulation, with optimal value $p = 0.004$}. We obtained a description of the algorithm from the authors via personal communication. Our implementation reproduces the penalty weight $p = 0.004$ reported in DD23 when applied to DD23's fingerprint time series data, confirming a faithful replication of their method.


To define a data generating process from \eqref{eq: DD23}, we need to set the parameters $t_0, \alpha_0, \mu_0, \sigma^2, A, \tau_r$, where $\tau_r = t_c - t_0$, $\lambda_0 = -\alpha_0^2/(4A)$, $m = \mu_0 - \alpha_0/(2A)$ following the SI and provided code of DD23. For $t_0, \alpha_0, \mu_0, \sigma^2$ we use the values estimated in DD23 from the observational data, i.e., $t_0 = 1924$, $\alpha_0 \approx 3.06$, $\mu_0 \approx 0.248$, $\sigma^2 \approx 0.296$. The other estimated parameters in DD23 are $A \approx 0.856$ and $\tau_r \approx 141$. Aside from this parameter combination, we evaluate $(A, \tau_r) \in \{(0.4, 250), (0.05, 2000)\}$.
For a given parameter vector, we sample a time series of simulated observations from Equation \eqref{eq: DD23}. Then we fix the estimate of $t_0$ to the simulation's true value, $1924$ (in DD23 this value is determined by their first method, the moment estimator), and follow the second method of DD23: We use the MLE without penalty for an initial parameter estimate, and cross-validation on 1000 time series simulated with this parameter vector to determine the penalty weight $p$. We then run the penalized MLE and the subsequent parametric bootstrap procedure as described in DD23, yielding one realization of the CI. We repeat the entire procedure 1000 times for each of the three parameter vectors described above. From these 1000 realizations of the CI, we approximate its distribution and arrive at the numbers reported above and in Figure \ref{fig:coverageSimulation}.

In the publicly available part of the code used in DD23, a general purpose optimizer is used to obtain the MLE. To ensure $A>0$ in the likelihood, the authors transform the input argument of the implemented likelihood function via $A \leftarrow \max(0.1, A)$, i.e., all inputs $A<0.1$ are treated as $0.1$. If the likelihood is indeed maximized at a value $A<0.1$, the optimizer therefore returns $A=0.1$ (to be more precise, the return value is the first value of $A$ that is tested and fulfills $A \leq 0.1$, which typically is a value just below $0.1$). The reason behind this functionality seems to be the requirement that $A>0$. But effectively truncating the range of possible values of $A$ from below by $0.1$ can introduce a severe bias that also strongly influences the estimation of $t_c$ as shown in Figure \ref{fig:coverageSimulation}. We therefore provide an alternative implementation that enforces $A>0$ by reparametrizing it as $A^\prime = \log(A)$, optimizing the likelihood over $A^\prime \in \mathbb{R}$, and transforming back via $A = \exp(A^\prime)$. Since the MLE is invariant under monotone reparametrization, this yields identical results (up to numerical precision) when the artificial barrier $A \geq 0.1$ is not triggered and is more faithful to the theoretical estimator in DD23 otherwise. See Figure \ref{fig:coverageSimulation} for a comparison of the bootstrapped CIs from both implementations.

We verified with the authors of DD23 that our interpretation of the CI procedure aligns with their intended meaning---in particular, that the coverage requirements for the CI are independent of the EWS test (significant increase of variance and autocorrelation). Additional checks that incorporate a dependence on the EWS test also reveal a failure in coverage (not shown here). Similarly, repeating the above procedure with the penalty weight fixed at the value used in DD23, namely $p=0.004$, also results in a failure in coverage. Finally, one could also subject the data-driven choice of $t_0$ to evaluation; however, fixing it to its true simulation value, as we do in our simulation study, should only improve the precision of the MLE and the CI.

\subsection*{Bayesian estimation of tipping times}

We adopt a Bayesian interpretation of DD23's model by assigning explicit priors to all free parameters in Eq. \eqref{eq: DD23} as well as the Ornstein-Uhlenbeck (OU) process used to model the stationary period. In order to fully propagate the uncertainty from all model parameters, we combine their two-step inference procedure into a single joint probability model over the stationary $X_{t \leq t_0}$ and nonstationary $X_{t > t_0}$ transitions,
\begin{align*}
    X_{t > t_0} &\sim \text{TP}(A, \tau_r, \alpha, \mu_t, \sigma^2)\,,\\
    X_{t \leq t_0} &\sim \mathcal{N}( X_{t-1}\rho + \mu_0 (1 - \rho), \gamma^2 (1 - \rho^2))\,,\\
    \alpha &\sim \mathcal{U}(0, 10)\,,\\
    \mu_0 &\sim \mathcal{U}(-10, 10)\,,\\
    \sigma_0 &\sim \mathcal{U}(0, 10)\,,\\
    \tau_r &\sim \mathcal{U}(0, 10^4)\,,
\end{align*}
where $\gamma^2 = \sigma_0^2 / 2\alpha$ is the conditional variance, $\rho = \exp(-\alpha \Delta)$ is the autocorrelation, and $\Delta$ is the time step size ($\frac{1}{12}$ years). As noted by DD23, the transition density for the non-stationary part, $\text{TP}$, is not available in closed form and is thus approximated using the Strang splitting scheme as implemented by DD23 (see corrected supplement of \cite{Ditlevsen2023WarningCirculation}). The combination of the stationary and non-stationary segments into a single model is justified by observing that using only point estimates of $\alpha$, $\sigma_0$, and $\mu_0$, as done by DD23, neglects uncertainty in the estimate of the OU parameters. We assign noninformative uniform priors to all other model parameters except for $A$. The parameter ranges for the uniform priors are chosen to provide numerical stability while avoiding influence on the resulting posterior. Note that the use of a double-bounded uniform prior for the ramping parameter assumes that tipping will occur within the next 10,000 years; this reflects an assumption of DD23's model, i.e. that tipping time must occur within a finite time span. For the parameter, $A$, we consider three cases: first, we apply the same improper penalty term described by DD23, $\log p(A) = -0.004\, n(1 / A - 1)$ with $n$ equal to the number of time steps in the nonstationary period. Second, we consider the case where the prior is set to be $A \sim \mathcal{U}(0.1,10)$ and exclude the penalty term. In both of these cases, we use the original implementation provided by DD23 which arbitrarily clips values of $A < 0.1$. Finally, we consider an alternative prior $\log A \sim \mathcal{N}(0,2)$, which applies a similar but more modest penalty to small (and large) values of $A$ and assigns zero density to $A = 0$ using the corrected implementation of the likelihood described in the previous section. For each model configuration, we draw 10,000 approximate samples from the posterior distributions conditioned on the original fingerprint published by DD23 using the Differential Evolution Markov Chain Monte Carlo (DE-MCMC) sampler of \cite{terbraakDifferentialEvolutionMarkov2008} as implemented by the \verb|BayesianTools| R library \cite{hartigBayesianToolsGeneralPurposeMCMC2017}. An additional 90,000 initial samples are simulated and discarded as burn-in to ensure convergence of the Markov chains, which is assessed via standard diagnostics (Gelman-Rubin $\hat{R} < 1.1$ and effective sample size $> 1000$). 

When using DD23's original choice of improper prior, we obtain a posterior credible interval for $t_c$ of 2039-2109 with median $t_c = 2065$, which is similar to their bootstrap confidence interval. This should be expected due to the strong influence of the penalty term. Under the uniform prior with $A > 0.1$, we obtain a posterior credible interval of 2050-2350 with median $t_c = 2110$. Results for the alternative prior using the corrected implementation of the likelihood (which should be regarded as the most reliable) are reported in the main text. We also tried running MCMC on DD23's original code without clipping (i.e. $A \sim \mathcal{U}(0,10)$); however, the Markov chains failed to converge within 100,000 iterations ($\hat{R} > 1.1$).

\subsection*{Corrections to DD23's fingerprint calculation}

An important practical point, in addition to their specific choice of fingerprint, is that DD23 make two numerical errors in the construction of the fingerprint time series. First, in their latitudinal weighting during area-averaging of SST in the SPG region, they multiply the SST values by the gridcell weights but do not divide by the sum of the weights (i.e. $\sum wx$ instead of $\sum wx/\sum w$). Additionally, instead of removing sea ice gridcells from the HadISST1 data, they set the value of all sea ice gridcells to -1.8 $^{\circ}$C, and include these datapoints in the SPG and global SST means. As the relationship between ocean heat transport (OHT) and sea ice is qualitatively different from that between OHT and SST, there is no fixed temperature value that can include sea ice in an AMOC fingerprint based on SSTs. Thus, when constructing our fingerprints from HadISST1 data we do not include sea ice cells. While correcting for these inconsistencies does not substantially alter the estimated tipping times, the increases in variance and autocorrelation become statistically insignificant according to the test in DD23.

\subsection*{MOC$_z$ data}
Bryden et al. 2005 \cite{bryden2005Slowing} estimated the strength of the overturning at 26N using five hydrographic sections from October 1957, August-Sep 1981, July-August 1992, February 1998, and April 2004. Following DD23, we call this set of datapoints the "MOC$_z$" data (as opposed to MOC$_z$ generally representing the strength of the AMOC in depth space). Since the months of these observations introduce a seasonal aliasing to the data \cite{Frajka-Williams2019AtlanticVariability}, Kanzow et al. 2010 \cite{Kanzow2010Seasonal} corrected the original MOCz data by subtracting the seasonal anomalies of the upper mid-ocean transport (T$_\text{UMO}$), which they calculated from 2004-2008 RAPID-MOCHA \cite{RAPID} data. However, updated observations of T$_\text{UMO}$ seasonal cycle reveal a more uneven cycle than that used by \cite{Kanzow2010Seasonal}, with a pronounced decrease in transport in Oct-Dec \cite{Perez-Hernandez2023TheGyre}. Following \cite{Kanzow2010Seasonal}, we calculate the T$_\text{UMO}$ seasonal anomalies from 2004-2024 RAPID-MOCHA data, and subtract the corresponding values from the original MOC$_z$, resulting in a different AMOC reproduction.

As DD23 have not published the code for calculating their fingerprint optimization or for plotting their Figure 7, we have approximated their process as described in the main text, by (i) subtracting the mean of the MOC$_z$ data, (ii) dividing the data by 3.8 Sv/K \cite{Caesar2018AMOCFingerprint}, (iii) shifting the result down by around 0.3 K. While this shifted data is closer to the SST fingerprints, when those fingerprints are correctly calculated, both versions of the MOC$_z$ data are a better fit to the SPG-1xGMT fingerprint than to SPG-2xGMT. 

\subsection*{Alternative fingerprints and datasets}
We extended the analysis beyond the HadISST1 dataset used in DD23 to include two additional observational products: ERSSTv5 \cite{Huang2017ExtendedIntercomparisons} and HadCRUT5 \cite{Morice2021AnSet}. For each dataset, we computed three distinct fingerprints:
\begin{enumerate}
    \item The index defined in DD23 (SPG SSTs minus twice the global mean SST). 
    \item The classical SPG-based AMOC index (SPG SSTs minus the global mean SST) \cite{Caesar2018ObservedCirculation}.
    \item The dipole fingerprint, calculated as the difference between averaged SSTs in the North Atlantic ($45^{\circ}$--$80^{\circ}$N, $70^{\circ}$W--$30^{\circ}$E) and the South Atlantic ($0^{\circ}$--$45^{\circ}$S, $70^{\circ}$W--$30^{\circ}$E) \cite{Roberts2013ACirculation, Jackson2020FingerprintsAMOC}.
\end{enumerate}

Additionally, we investigated the propagation of observational uncertainty by considering a 200-member uncertainty ensemble of HadCRUT5 \cite{Morice2021AnSet}. The tipping time $t_c$ was estimated by maximizing the likelihood of the observed discrete-time increments under the statistical fold-bifurcation model of DD23. 
For the HadCRUT5 dataset, observational uncertainty was propagated by applying the MLE procedure independently to each of the 200 ensemble members. 

To ensure comparability with DD23, we used the same time-interval settings (starting in 1870, with forcing initiated in 1924). We determined an optimal penalization term for each of the observation time series as described by DD23 in personal correspondence. This uniform treatment enables a direct assessment of how sensitive the tipping estimate is to choices in data processing and fingerprint construction. 
For transparency, we did not pre-select time series based on significant trends in conventional early-warning indicators. 
Indeed, none of the observational fingerprints considered here display a statistically significant increase in variance and autocorrelation at the 5\% level according to DD23's criterion for avoiding spurious tipping time predictions mentioned above. 
While such a signal was reported in DD23 for their preferred fingerprint, we find that it disappears once the errors in the surface-area weighting and sea-ice handling are corrected (see above).
\end{document}